\documentclass[paper, 12pt, letterpaper]{JHEP} \usepackage{amsmath}
\usepackage{amssymb} \usepackage{graphics} \usepackage{epsfig}

 \def\C{{ \mathbb C}} \def\R{{ \mathbb R}} \def\Z{{
    \mathbb Z}} \def\H{{ \mathbb H}}  \def\P{{
    \mathbb P}}
    \def\tr{\operatorname{tr}}

 \def\tr{{\rm tr\, }} \def\dim{{\rm dim}}

\def\id{\protect{{1 \kern-.28em {\rm l}}}}

\newcommand{\be}{\begin{equation}} \newcommand{\ee}{\end{equation}}
\newcommand{\bea}{\begin{eqnarray}} \newcommand{\eea}{\end{eqnarray}}
\newcommand{\beann}{\begin{eqnarray*}}
  \newcommand{\eeann}{\end{eqnarray*}}
\newcommand{\bfig}{\begin{figure}} \newcommand{\efig}{\end{figure}}
\newcommand{\nn}{\nonumber}
\newcommand{\ba}{\begin{array}}\newcommand{\ea}{\end{array}}

\newtheorem{Proposition}{Proposition}[section]

\newtheorem{Theorem}{Theorem}[section]
\newtheorem{Lemma}{Lemma}[section]
\newtheorem{Corrolary}{Corrolary}[section]

\newcommand{\bp}{\begin{Proposition}}
  \newcommand{\ep}{\end{Proposition}}
\newcommand{\bt}{\begin{Theorem}} \newcommand{\et}{\end{Theorem}}
\newcommand{\bl}{\begin{Lemma}} \newcommand{\el}{\end{Lemma}}
\newcommand{\bc}{\begin{Corrolary}} \newcommand{\ec}{\end{Corrolary}}

   \def\ep{\eps}
      \def\cn{{\cal N}}

\author{K. Landsteiner\\
  Instituto de F{\'\i}sica Te\'orica C-XVI\\
  Universidad Aut{\'o}noma de Madrid\\
  28049 Madrid,Spain\\Karl.Landsteiner@uam.es}

\author{C. I. Lazaroiu\\
  Humboldt Universit\"at zu Berlin\\
  Newtonstrasse 15, 12489 Berlin-Adlershof, Germany\\
  calin@physik.hu-berlin.de}

\author{Radu Tatar\\
  Theoretical Physics Group \\
  Lawrence Berkeley National Laboratory\\
  Berkeley, CA 94720, USA\\rtatar@socrates.berkeley.edu}

\title{Chiral field theories from conifolds} \abstract{ We discuss the
  geometric engineering and large $N$ transition for an $\cn=1$ $U(N)$
  chiral gauge theory with one adjoint, one conjugate symmetric, one
  antisymmetric and eight fundamental chiral multiplets.  Our IIB
  realization involves an orientifold of a non-compact Calabi-Yau
  $A_2$ fibration, together with D5-branes wrapping the exceptional
  curves of its resolution as well as the orientifold fixed locus. We
  give a detailed discussion of this background and of its relation to
  the Hanany-Witten realization of the same theory. In particular, we
  argue that the T-duality relating the two constructions maps the
  $\Z_2$ orientifold of the Hanany-Witten realization into a $\Z_4$
  orientifold in type IIB. We also discuss the related engineering of
  theories with $SO/Sp$ gauge groups and symmetric or antisymmetric
  matter.}

\preprint{HU-EP-03/39 \\
  IFT-UAM/CSIC-03-37 \\
  LBNL-53843,~UCB-PTH-03/24}

\begin{document}

\tableofcontents

\pagebreak

\vskip .6in

\section{Introduction}
\label{intro}
String theory provides nontrivial information about supersymmetric
gauge theories by means of geometric engineering.  An interesting
class of models is based on local conifold geometries \cite{vafa,civ,
  Cachazo_Vafa, Cachazo_Vafa_more}.  These arise in IIB string theory
on a resolved ADE fibration over the complex plane, with D5-branes
partially wrapped on the exceptional divisors. On the common
non-compact part of the world volumes, one obtains an $\cn=1$
supersymmetric field theory in four dimensions, whose precise nature
is dictated by the Calabi-Yau geometry. The gauge group of such models
is a product of $U(N)$ factors, with $N$ fixed by the number of
partially wrapped D5-branes. It has been argued that at large $N$ but
fixed $g_{string}N$ such models undergo a transition in which the
resolved geometry is replaced by the background obtained through a
conifold transition, realized by blowing down all exceptional curves
and smoothing out the resulting singular fibration. The exact
effective superpotential of the gauge theory can then be calculated by
evaluating period integrals of the deformed geometry.

This construction leads to the conjecture that the effective
superpotential of a confining $\cn=1$ supersymmetric gauge theory can
be computed using a certain holomorphic \cite{holo} matrix model
\cite{DV,DV2,DV3,holo}.  In a related development it was shown that
the loop equations of this model can be recovered by studying certain
generalized Konishi anomalies of the gauge theory \cite{cdsw}.

The approach described above leads to theories whose gauge group is a
product of $U(N)$ factors, with matter in adjoint or bifundamental
chiral multiplets. To engineer more general models, one can consider
orientifolds of such backgrounds, possibly supplemented by the
introduction of further D-branes.  Such extensions of the framework of
\cite{vafa,civ, Cachazo_Vafa, Cachazo_Vafa_more} have been much less
studied. The simplest examples of $\Z_2$ orientifolds where considered
in \cite{Roemelsberger, eot,ookouchi}, while a considerably more
complicated system was recently analyzed in \cite{us, or}.

It is known that orientifolds in the presence of D-branes can lead to
chiral field theories under specific conditions. For IIA backgrounds,
the standard example is provided by the work of
\cite{Karl_chiral,hana,kutasov}, which leads to an ${\cal N}=1$ gauge
theory with net chirality through a combination of subtle phenomena.
More precisely, it was shown in \cite{Karl_chiral,hana,kutasov} that
one can obtain an ${\cal N}=1$ $U(N)$ supersymmetric model with one
adjoint chiral multiplet, eight fundamental multiplets as well as
multiplets transforming in the antisymmetric and conjugate symmetric
representations of the gauge group. The IIA realization of this model
involves a certain $\Z_2$ orientifold of a Hanany-Witten configuration
with three NS5-branes and N pairs of D4-branes, with the addition of
half D6-branes along the O6 plane, taken such that they end on the
central NS5-brane \footnote{A IIB realization of this system was
  discussed in \cite{park1}. This is obtained by performing a certain
  T-duality which differs from the one we shall consider in this
  paper.}.  The Konishi anomalies of this chiral theory were recently
studied in \cite{chiral1}, where it was shown that they are reproduced
by a certain holomorphic matrix model. In the present paper, we use
the methods of \cite{OT} to propose a novel IIB realization of this
theory as a $\Z_4$ orientifold of a certain non-compact Calabi-Yau
$A_2$ fibration with D5-branes wrapping the exceptional fibers of the
resolution, and supplemented by the introduction of eight fractional
D-branes lying along the orientifold fixed locus (the latter branes
are fractional with respect to the $\Z_2$ orbifold group of the
orientifold group).

The paper is organized as follows.  In section \ref{ft} we introduce
our field-theoretic model.  In section \ref{ge} we give a detailed
construction of the relevant IIB background.  In particular, we
present global and local models of the resolved geometry following
\cite{OT} as well as an explicit construction of the T-duality which
maps this background to the Hanany-Witten construction.  Our proposal
for the T-duality map follows an ansatz first considered in \cite{or}
and predicts that the $\Z_2$ orientifold of \cite{Karl_chiral} is
mapped to a $\Z_4$ orientifold in IIB.  In section \ref{gt} we
consider the large N transition and effective superpotential.  Section
\ref{so_eng} discusses the geometric engineering of the related
theories with $SO/Sp$ gauge groups and symmetric/antisymmetric matter.
Section 6 contains our conclusions.  In the appendices we discuss the
relation to general log-normalizable Calabi-Yau $A_2$ fibrations, a
fractional brane construction related to our engineering and the
orientifold projection on the Chan-Paton factors.

\section{The field theory model}
\label{ft}
Our ${\cal N}=1$ model contains chiral multiplets $\Phi, A, S$ in the
adjoint, antisymmetric and conjugate symmetric representations of
$U(N)$ as well as eight quarks $Q_1\dots Q_8$ in the fundamental
representation. We consider the tree-level superpotential:
\begin{equation}
\label{W_tree}
W_{tree}=\tr \left[ W(\Phi) + S \Phi A\right] + \sum_{f=1}^{8}{Q_f^T
  SQ_f}~~, \end{equation} where: \begin{equation}
\label{W}
W(z)=\sum_{j=1}^{d+1}{\frac{t_j}{j}z^j} \end{equation} is a complex polynomial of
degree $d+1$. The fields are constrained by $S^T=S, A^T=-A$ and the
gauge transformations take the form: \begin{equation}
\label{field_gauge}
\Phi\rightarrow U \Phi U^\dagger~~,~~S\rightarrow {\bar U} S
U^\dagger~~,~~A\rightarrow U A U^T~~,~~Q_f\rightarrow UQ_f~~, \end{equation}
where $U$ is valued in $U(N)$. The restriction to eight quark flavors
follows from cancellation of the chiral anomaly. This model was studied in \cite{chiral1}
through the method of generalized Konishi anomalies, which allows
for a proof of the Dijkgraaf-Vafa correspondence in this situation.

As discussed in \cite{Karl_chiral,hana,kutasov}, our model can be
obtained in IIA string theory through a Hanany-Witten configuration
involving a $\Z_2$ orientifold of a system of NS5, D4 and half
D6-branes (this construction is recalled -- and slightly extended-- in
Subsection \ref{hanany_witten}).  Our first purpose is to determine
the geometric engineering of this field theory and extract its
relation with the Hanany-Witten construction.  We shall find that the
engineering of our system is quite nontrivial, and in particular it
involves $\Z_4$ orientifolds of resolved Calabi-Yau $A_2$ fibrations.
Such backgrounds contain D5-branes wrapped over the exceptional fibers
of the resolution, as well as D5-branes stretching along the fixed
point set of the orientifold 5-plane\footnote{In contrast the
  T-duality considered in \cite{park1,park2} results in a $\Z_2$
  orientifold 7-plane together with a collection of D3-branes and
  D7-branes.}.

\section{Chiral field theories from geometric engineering}
\label{ge}

In this section we engineer our model by considering a $\Z_4$
orientifold of a IIB background with D5-branes.  We also relate this
to the Hanany-Witten description of \cite{Karl_chiral,hana,kutasov} by
applying T-duality with respect to a certain $U(1)$ action and using
an ansatz for the dual coordinates originally considered in \cite{or}.

\subsection{The IIB background}

Our starting point is a Calabi-Yau $A_2$ fibration, chosen to admit a
certain holomorphic $\Z_4$ action. We shall first consider the
singular limit of this fibration, then provide global and local models
of its resolution following \cite{OT}.

\subsubsection{The singular limit $X_0$}

Consider the Calabi-Yau hypersurface $X_0$ defined by the equation:
\begin{equation}
\label{X0}
xy=u(u-W'(z))(u+W'(z))~~, \end{equation} where $z,u,x,y$ are the affine
coordinates of $\C^4$ and $W$ is a polynomial of degree $d+1$ with
complex coefficients.  Throughout this section, we restrict to the
generic case when the derivative $W'$ has only simple roots, which we
denote by $z_j$. The index $j$ runs from $1$ to $d$.

\begin{figure}[hbtp]
\begin{center}
  \scalebox{0.7}{\input{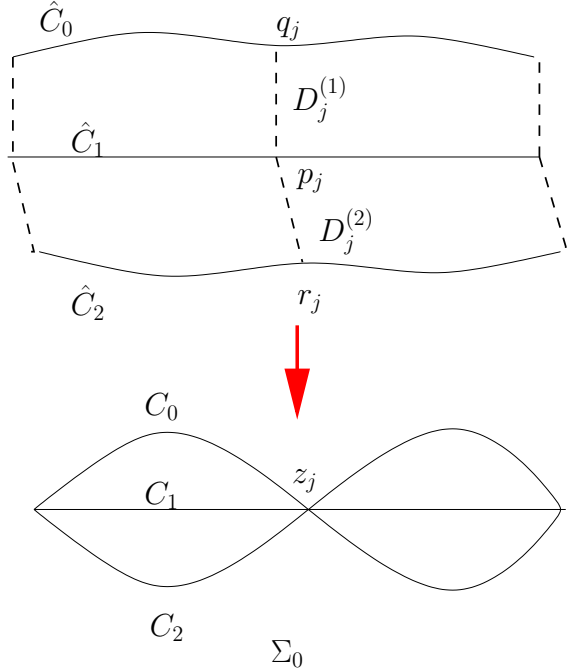}}
\end{center}
\caption{The degenerate curve $\Sigma_0$ and its transform ${\hat
    \Sigma}$ (to be discussed in the next subsection). The orientifold
  action fixes ${\hat C}_1$ and flips ${\hat C}_0$ and ${\hat C}_2$.}
\label{curve}
\end{figure}

The space (\ref{X0}) is a singular $A_2$ fibration over the $z$-plane,
which admits a 3-section $\Sigma_0$ obtained by requiring $x=y=0$.
Together with (\ref{X0}) this gives the planar curve with equation:
\begin{equation}
\label{Sigma_0}
u(u-W'(z))(u+W'(z))=0~~.  \end{equation} Thus $\Sigma_0$ has three rational
components, which we index as follows: \begin{equation}
\label{Cj}
C_0:~u=-W'(z)~~,~~C_1:~u=0~~,~~C_2:~u=+W'(z)~~.  \end{equation} One easily checks
that the components meet only when $u=0$ and $z$ coincides with a
critical point of $W$. In fact $\Sigma_0$ has triple points sitting at $u=0~,~z=z_j$
and no other singular points (see figure \ref{curve}).

The total space (\ref{X0}) also has $d$ singular points, which
coincide with the triple points of $\Sigma_0$. The fiber sitting above
$z_j$ is a standard $A_2$ singularity: \begin{equation}
\label{A2sing}
xy=u^3~~.  \end{equation} When $z$ is away from all $z_j$, the fiber $X_0(z)$ is
a smooth deformation of (\ref{A2sing}).  Resolving each singularity of
$X_0$ leads to a smooth space ${\hat X}$ which we describe in the next
subsection.

Below, we shall be interested in the $U(1)$ action on $X_0$ given by:
\begin{equation}
\label{U0}
\rho_0(\theta):~~(z,u,x,y)\rightarrow (z,u,e^{i\theta}x,
e^{-i\theta}y)~~, \end{equation} whose fixed point set coincides with the
multisection $\Sigma_0$. We also consider the bi-holomorphism: \begin{equation}
\label{or0}
\kappa_0:~~(z,u,x,y)\longrightarrow (z,-u,-y,x)~~, \end{equation} whose fixed
point set $O_0$ coincides with the central component $C_1$. We shall use the later to define
a $\Z_4$ orientifold of our background, as explained in more detail
below.  The square of the generator (\ref{or0}) is given by: \begin{equation}
\kappa_0^2:~~(z,u,x,y)\longrightarrow (z,u,-x,-y)~~ \end{equation} and thus
$\kappa_0^4=id$.

\paragraph{Observation:} The singular space (\ref{X0}) can be obtained by
starting with a generic Calabi-Yau $A_2$ fibration and requiring
invariance under (\ref{or0}).  This is explained in Appendix A.

\subsubsection{The resolution}

Consider the space ${\hat X}$ obtained by resolving each of the
singular $A_2$ fibers $X_0(z_j)$.  Following \cite{OT}, we describe it
as the complete intersection in $\P^1\times \P^1\times \C^4$ cut by
the equations: \begin{eqnarray}
\label{hatX}
\beta_1 (u+W'(z))&=&\alpha_1 x\nn\\
\alpha_2(u-W'(z))&=&\beta_2 y\\
\alpha_1 \beta_2 u&=&\beta_1\alpha_2~~.\nn\\
u(u-W'(z))(u+W'(z))&=&xy\nn \end{eqnarray} We let $[\alpha_i,\beta_i]$ be the
homogeneous coordinates on the two $\P^1$ factors ($i=1,2$).
Forgetting these gives the resolution map $\tau:{\hat X}\rightarrow
X_0$.  The exceptional fibers of $\tau$ are given by $z=z_j,u=x=y=0$
and $\alpha_2=0$ or $\beta_1=0$. The case $\alpha_2=0$ defines a
rational curve $D_j^{(1)}$ which can be identified with
$\P^1[\alpha_1,\beta_1]$, while $\beta_1=0$ gives the curve
$D_j^{(2)}$, identified with $\P^1[\alpha_2,\beta_2]$. The rational
curves $D_j^{(1)}$ and $D_j^{(2)}$ intersect transversely at a point
$p_j$ which sits at $z=z_j,u=x=y=\alpha_2=\beta_1=0$ (see figure
\ref{curve}) . If one views ${\hat X}$ as a fibration over the
$z$-plane, then its fiber ${\hat X}(z)$ can be identified with
$X_0(z)$ when $W'(z)\neq 0$, and coincides with the minimal resolution
of $X_0(z_j)$ when $z=z_j$.

Our next task is to lift the circle action (\ref{U0}) and $\Z_4$
generator (\ref{or0}) to the resolved space. It is easy to see that
(\ref{U0}) is covered by the following $U(1)$ action on ${\hat X}$:
\begin{equation}
\label{Uhat}
{\hat
  \rho}(\theta):~~(z,u,x,y,[\alpha_1,\beta_1],[\alpha_2,\beta_2])\longrightarrow
(z,u,e^{i\theta}x,e^{-i\theta}y,[e^{-i\theta}\alpha_1,\beta_1],[\alpha_2,e^{i\theta}\beta_2])
\end{equation} while (\ref{or0}) lifts to: \be
\label{orhat}
{\hat
  \kappa}:~~(z,u,x,y,[\alpha_1,\beta_1],[\alpha_2,\beta_2])\longrightarrow
(z,-u,-y,x,[\beta_2,\alpha_2], [-\beta_1,\alpha_1])~~.  \end{equation}

\begin{figure}[hbtp]
\begin{center}
  \scalebox{0.7}{\input{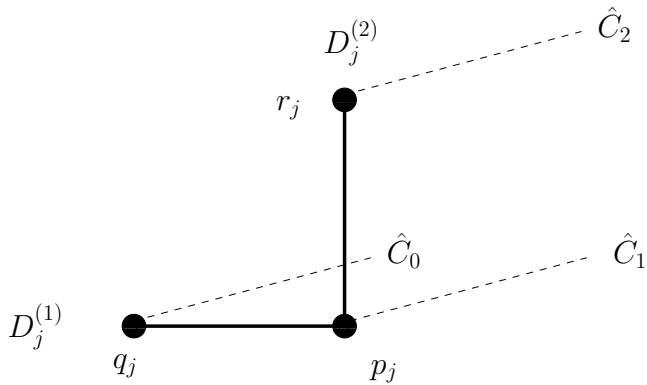}}
\end{center}
\caption{Intersection points of ${\hat C}_j$ with the resolved fiber
  ${\hat X}(z_j)$. We introduce D5-branes partially wrapped on the
  exceptional curves $D_j^{(1)}$ and $D_j^{(2)}$ and an orientifold
  5-plane whose internal part stretches along ${\hat C}_1$. For
  consistency, we also have eight (fractional) D5-branes stretched
  along the orientifold locus. These D-branes are fractional because
  the square of the orientifold generator is nontrivial and induces a
  $\Z_2$ orbifold action.}
\label{res_fiber}
\end{figure}

The fixed point set of (\ref{Uhat}) has three rational components
${\hat C}_0$, ${\hat C}_1$ and ${\hat C}_2$ which project to the
curves $C_0,C_1$ and $C_2$ via the resolution map.  These are given by
$x=y=0$ and: \begin{eqnarray}
  &{\hat C}_0&:~u+W'(z)=\alpha_1=\alpha_2=0\nn\\
  &{\hat C}_1&:~u=\alpha_2=\beta_1=0\\
  &{\hat C}_2&:~u-W'(z)=\beta_1=\beta_2=0~~.\nn \end{eqnarray} The
combination of these curves gives a 3-section ${\hat \Sigma}$ of the
fibration ${\hat X}$.  When $z$ coincides with a critical point $z_j$,
the curves ${\hat C}_1, {\hat C}_2, {\hat C}_3$ meet the resolved
fiber ${\hat X}(z_j)$ at three distinct points $q_j, p_j, r_j$ given
by $x=y=0$ and $\alpha_1=\alpha_2=0$, $\alpha_2=\beta_1=0$ and
$\beta_1=\beta_2=0$ respectively. The second of these is the
intersection of $D_j^{(1)}$ with $D_j^{(2)}$, while the first and last
point lie on these exceptional curves respectively (figure
\ref{res_fiber}). As expected, the curves ${\hat C}_j$ are separated
inside ${\hat X}(z_j)$, unlike their projections $C_j$ which meet at
$x=y=u=0~,~z=z_j$.

The $\Z_4$ generator (\ref{orhat}) flips ${\hat C}_0$ and ${\hat C}_2$
while fixing ${\hat C}_1$ --- in fact, its fixed point set ${\hat O}$
coincides with the central component ${\hat C}_1$. This action
exchanges $D_j^{(1)}$ and $D_j^{(2)}$. As mentioned above, we use
(\ref{orhat}) to define a $\Z_4$ orientifold of our IIB background. In
particular, we have an orientifold 5-plane with worldvolume given by
$\R^{1,3}\times {\hat C}_1$, which we shall take to be an O5$^-$ plane
for agreement with the Hanany-Witten construction discussed below.

One can check that this D-brane configuration reproduces the field
content of our chiral theory.  One way to establish this is to
consider the limit $W'\equiv 0$, with the exceptional $\P^1$'s blown
down. Then the background becomes a $\Z_4$ orientifold of a trivial
$\C^2/\Z_3$ fibration over the complex plane.  The eight D5 branes
along the orientifold fixed locus live on the zero section of this
fibration, while the D5-branes wrapped over the exceptional $\P^1$'s
are represented as D3-branes transverse to the total space of the
fibration and delocalized along $z$. This allows one to implement the
orientifold action by considering a projection originally used in
\cite{park1} (albeit in a different context).  The details of this
analysis can be found in Appendix \ref{fractional}. An alternate
derivation of the massless spectrum (which does not involve taking the
fractional brane limit) is given in Appendix C. In this second
approach, the symmetric and antisymmetric multiplets arise from
strings stretching between the D5-branes wrapping the exceptional
fibers, while the fundamentals arise from strings stretching between
these branes and the eight fractional D5 branes wrapped over ${\hat
  C}_1$ (the massless states of all these types are localized at the
intersection points $p_j$). Finally, the adjoint field arise from
strings stretching between the D5 branes wrapping the exceptional
fibers $D_j^{(\alpha)}$. Their massless components correspond to
strings stretching between branes wrapping the same exceptional curve,
which give rise to massless states living on that curve.

\subsubsection{A local model of the resolution}

As in \cite{OT}, we can consider a local model ${\tilde X}\subset
{\hat X}$ of the resolution, which is obtained by gluing three copies
${\cal U}_j$ ($j=0\dots 2$) of $\C^3$ (with affine coordinates $x_j,
u_j, z_j$) according to the identifications: \begin{eqnarray}
  (z_1, x_1, u_1)&=&(z_0,~~\frac{1}{u_0},~~x_0 u_0^2-W'(z_0)u_0)\nn\\
  (z_2, x_2, u_2)&=&(z_1,~~\frac{1}{u_1},~~x_1 u_1^2-W'(z_1)u_1)~~.
\end{eqnarray} The restricted projection $\tau:{\tilde X}\rightarrow X$ is given
by: \begin{eqnarray}
  (z,u,x,y)&=&(z_0,~~x_0u_0-W'(z_0),~~x_0,~~u_0[x_0u_0-W'(z_0)][x_0u_0-2W'(z_0)])\nn\\
  (z, u, x,
  y)&=&(z_1,~~x_1u_1,~~x_1[x_1u_1+W'(z_1)],~~u_1[x_1u_1-W'(z_1)])\\
  (z, u, x,
  y)&=&(z_2,~~x_2u_2+W'(z_2),~~x_2[x_2u_2+W'(z_2)][x_2u_2+2W'(z_2)],~~u_2)\nn
\end{eqnarray} In this description, the exceptional curve $D_j^{(1)}$ sits at
$z_1=z_j~,~u_1=0$, while $D^{(2)}_j$ is given by $z_1=z_j~,~x_1=0$.

The orientifold action (\ref{orhat}) takes the form:
\begin{equation}
\label{orlocal}
{\hat \kappa}:~~(z_1,x_1,u_1)\rightarrow (z_1,u_1, -x_1)
\end{equation}
with fixed point set ${\hat O}$ given by ${\hat C}_1$.

As in \cite{OT}, the brane construction of \cite{Karl_chiral} arises
by performing T-duality with respect to the $U(1)$ action (\ref{Uhat})
on ${\hat X}$, which has the following form in local coordinates:
\begin{equation}
\label{U1action}
{\hat \rho}(\theta):~~(z_j, x_j, u_j)\longrightarrow
(z_j,~~e^{i\theta}x_j,~~e^{-i\theta}u_j)~~.  \end{equation} The fixed point locus
of (\ref{U1action}) consists of the three components ${\hat C}_j$,
with local equations: \begin{eqnarray} {\hat C}_j:~~u_j=x_j=0~~, \end{eqnarray} which take
the following form in the coordinates of the patch ${\cal U}_1$:
\begin{eqnarray}
\label{hatC_local}
{\hat C}_0&:&~~x_1u_1=-W'(z)~~,~~x_1=\infty\nn\\
{\hat C}_1&:&~~x_1=u_1=0\\
{\hat C}_2&:&~~x_1u_1=+W'(z)~~,~~u_1=\infty\nn~~.  \end{eqnarray} The $U(1)$
action stabilizes the exceptional curves $D_j^{(\alpha)}$.

\subsection{Relation to the orientifolded Hanany-Witten construction}
\label{hanany_witten}

Under T-duality along the orbits of (\ref{U1action}), the loci ${\hat
  C}_j$ become three NS5-branes denoted by ${\cal N}_j$, while the
D5-branes wrapping $D_j^{(\alpha)}$ are mapped into two stacks of
$D_4$-branes stretching between them, which we denote by ${\cal
  D}_j^{(\alpha)}$.

To see the T-dual picture explicitly, let us follow \cite{or} by
combining $x_1$ and $u_1$ into the quaternion coordinate $X=x_1+{\bf
  j}u_1$ (where ${\bf j}$ is the second imaginary quaternion unit) and
notice that the $U(1)$ action (\ref{U1action}) becomes:
\begin{equation} X\longrightarrow e^{i\theta}X~~.
\end{equation} The hyperkahler moment map ${\vec
  \mu}:\H[X]=\C^2[x_1,u_1]\rightarrow \R^3$ of this action presents
$\C^2[x_1,u_1]$ as an $S^1$ fibration over $\R^3$ (the fiber collapses
to a point at the origin of $\R^3$).  Considering the real and complex
components of ${\vec \mu}$ allows us to introduce real coordinates
$(x^4\dots x^9)$ on the patch ${\cal U}_1$ by the equations:
\begin{eqnarray}
\label{coords}
x^4+ix^5&=&x_1u_1\nn\\
x^6&=&\frac{1}{2}(|x_1|^2-|u_1|^2)\nn\\
x^7&=&{\rm coordinate~on~the~}S^1{\rm~fiber~of~}{\vec \mu}~~.\\
x^8+ix^9&=&z_1\nn \end{eqnarray}

Then the dual NS5-branes are described by: \begin{eqnarray}
  {\cal N}_0&:&~~x^4+ix^5=-W'(z)~~,~~x^6=+\infty,~~x^7=0\nn\\
  {\cal N}_1&:&~~x^4=x^5=x^6=x^7=0\\
  {\cal N}_2&:&~~x^4+ix^5=+W'(z)~~,~~x^6=-\infty,~~x^7=0\nn~~,
\end{eqnarray} while the dual D4-branes ${\cal D}_j^{(1)}$ and ${\cal
  D}_j^{(2)}$ are localized at $x^8+ix^9=z_j$, $x^4=x^5=x^7=0$ and
extend in the direction $x^6$ between ${\cal N}_1$ and ${\cal N}_0$,
${\cal N}_2$ respectively (i.e. we have $x^6\geq 0$ for ${\cal
  D}_j^{(1)}$ and $x^6\leq 0$ for ${\cal D}_j^{(2)}$).

\begin{figure}[hbtp]
\begin{center}
  \scalebox{0.7}{\input{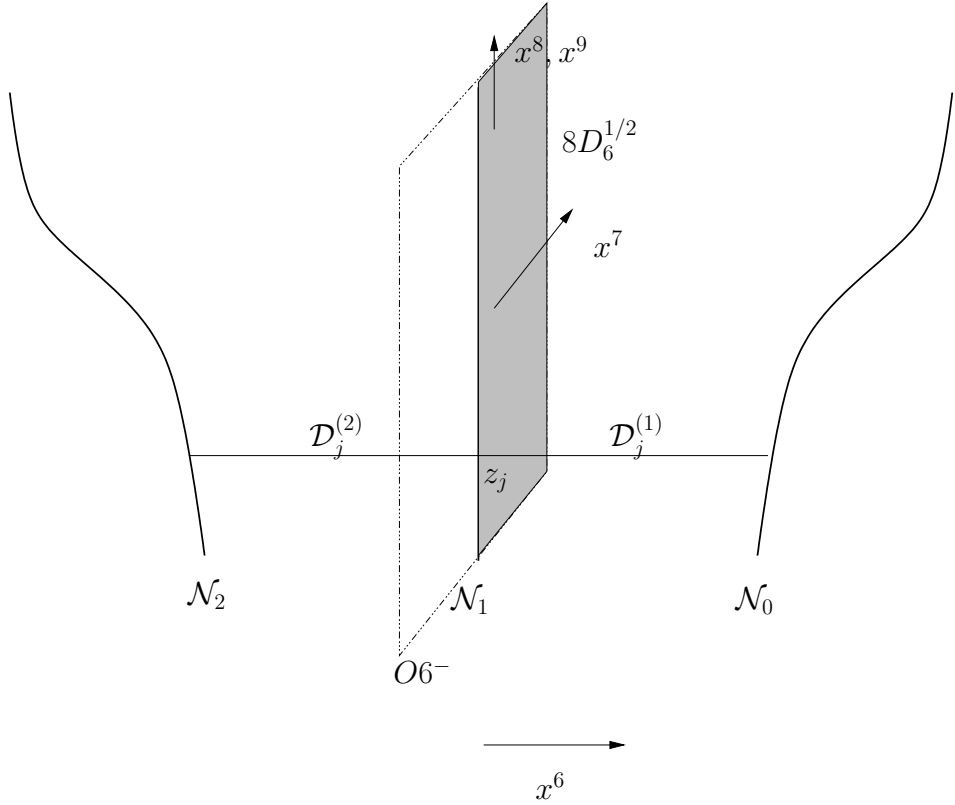}}
\end{center}
\caption{The T-dual brane configuration. }
\label{chiral_config}
\end{figure}

Equations (\ref{coords}) show that the orientifold action
(\ref{orlocal}) fixes $x^8$ and $x^9$ while changing the sign of
$x^4$, $x^5$ and $x^6$. The obvious relation: \begin{equation} {\hat
    \kappa}\circ{\hat \rho}(\theta)={\hat \rho}(-\theta)\circ {\hat
    \kappa} \end{equation} shows that the IIB orientifold action
reflects the coordinate $x^7$, which means that the IIA orientifold
action must fix this coordinate.  Hence T-duality produces an
orientifold 6-plane sitting at $x^4=x^5=x^6=0$, with action:
\begin{equation}
\label{ordual}
(x^4,x^5,x^6)\rightarrow (-x^4,-x^5,-x^6)~~ \end{equation} on the transverse
coordinates. Note that in the IIA construction we obtain a $\Z_2$
orientifold, even though we started with a $\Z_4$ orientifold in IIB.
This is because the first two of relations (\ref{coords}) are nonlinear.
Finally, our eight fractional D5-branes become eight half D6-branes stretching
along the O6 plane and
ending on the central NS5-brane (figure \ref{chiral_config}). This recovers the
Hanany-Witten configuration of
\cite{Karl_chiral}.

It is well-known \cite{Karl_chiral,hana,kutasov} that this
Hanany-Witten configuration produces the matter content of our chiral
field theory. More precisely, the effective theory lives on the
worldvolume of the D4-branes, which are identified in pairs by the
orientifold action, namely ${\cal D}_j^{(1)}\equiv {\cal D}_j^{(2)}$.
This theory has a $U(N)$ gauge group because the orientifold action
identifies gauge transformations along ${\cal D}_j^{(1)}$ and ${\cal
  D}_j^{(2)}$ through the relation: \begin{equation}
\label{U_or}
U_2=U_1^{-T}:=U~~.  \end{equation} The adjoint field $\Phi$ arises from strings
stretching between D4-branes of the same stack (i.e. between some
$D_j^{(\alpha)}$ and some $D_k^{(\alpha)}$). These give adjoint chiral
superfields $\Phi_\alpha$ on the worldvolumes of the two stacks, which
are identified by the orientifold projection as: \begin{equation}
\label{Phi_or}
\Phi_2=\Phi_1^T:=\Phi~~.  \end{equation} The symmetric and antisymmetric fields
$S,A$ arise from strings stretching from ${\cal D}_j^{(1)}$ to ${\cal
  D}_k^{(2)}$ and backwards.  The zero-modes of such strings give rise
to massless fields $\Phi_{12}$ and $\Phi_{21}$ in the bi-fundamental
representation of the unprojected gauge group $U(N)\times U(N)$. Then the
orientifold projection imposes: \begin{equation}
\Phi_{12}=-\Phi_{12}^T:=A~~{\rm~and~}~~\Phi_{21}=\Phi_{21}^T=S~~.  \end{equation}
The identification (\ref{U_or}) shows that $A$ and $S$ transform in
the antisymmetric and conjugate symmetric representations of $U(N)$.
Finally, the eight fundamentals $Q_f$ arise from strings
stretching between ${\cal D}_j^{(\alpha)}$ and the eight half D6-branes.

As explained in \cite{Karl_chiral}, this brane configuration also
produces a superpotential of type (\ref{W_tree}). Since we allow the
outer NS5-branes to be curved, we have a general polynomial
contribution $\tr W(\Phi)$ to the superpotential for the adjoint
field, where $W$ controls the shape of the outer NS5-branes (this is a
minor generalization with respect to \cite{Karl_chiral}, which
considered tilting but no bending of the NS5-branes ).

To check matching of RR-charges through T-duality, remember that the
pair (O6-plane, eight half D6-branes) in the Hanany-Witten realization
contains an orientifold plane which is divided in two halves by the
central NS5-brane. One half carries D6-brane charge $+4$, while the
other half carries D6-brane charge $-4$. This creates a jump in RR
charge density along the central NS5-brane which is compensated by the
eight half D6-branes.  In the geometric engineering construction this
corresponds to cancellation of twisted RR-tadpoles (in the twisted
sector of the $\Z_2$ orbifold subgroup of the $\Z_4$ orientifold
group). Notice that the ansatz (\ref{coords}) for the dual coordinates
predicts a single orientifold plane in the IIB dual --- this cannot be
divided into halves as in the Hanany-Witten description, because the
NS5 branes are eliminated when performing the T-duality. On the other
hand, it is clear that the T-duals of the half D6-branes cannot be
ordinary branes, but must be fractional in some sense \footnote{The D6 branes
span a half of the T-duality circle. This is reminiscent of the fractional
D3 branes at conifold singularity, where they become D4 branes on a
semicircle after a T-duality.}.  This is why
the IIB construction involves a further $\Z_2$ orbifold (implemented
by the $\Z_2$ subgroup of the orientifold group generated by ${\hat
\kappa}^2$). This $\Z_2$ orbifold appears due to taking the T-duality
orthogonal to the NS branes but along the O6 plane. 
This allows us to take the dual D5-branes to be
fractional with respect to this $\Z_2$ subgroup, which matches the
fact that the central D6 branes in the Hanany-Witten construction
carry only half of the usual RR charge of a D6 brane. It is remarkable
that the simple ansatz (\ref{coords}) (which was already tested in
\cite{or} in a different context) maps the $\Z_2$ orientifold of the
Hanany-Witten construction into a $\Z_4$ orientifold in IIB, thus
automatically implementing this fractionality requirement for the
central D5 branes.

Relation (\ref{orlocal}) gives the following action for the $\Z_2$
orbifold generator ${\hat \kappa}^2$: \be
\label{or_gen_local}
{\hat \kappa}^2:~~(z_1,x_1,u_1)\rightarrow (z_1,-x_1, -u_1)~~.  \ee
Using (\ref{hatC_local}), we find that the fixed locus of this action
coincides with ${\hat C}_1$.  Since our O5-plane sits along this
locus, we can view it as a fractional orientifold plane, i.e. a $\Z_2$
orientifold with fractional charge under the orbifold action
(\ref{or_gen_local}).  Indeed, one can resolve the $\Z_2$ orbifold
singularity induced by (\ref{or_gen_local}) by performing the blow-up
of ${\hat X}$ along the curve ${\hat C}_1$ (this amounts to blowing up
the point $p_j$ in every fiber ${\hat X}(z)$). Then our fractional
O5-plane becomes an O7-plane wrapping the locus $\P(N^*_{{\hat
    C}_1/{\hat X}})$, which is the $\P^1$ fibration over ${\hat C}_1$
`glued in' through this resolution.  To cancel the charge of this O7
plane, one must add eight D7 branes wrapping the same locus, which
correspond to our eight fractional D5-branes after blowing down to
recover the original space ${\hat X}$.  The strings stretched between
these fractional D5-branes and the D5-branes wrapping the exceptional
divisors ${\cal D}_j^{(\alpha)}$ give rise to the chiral multiplets in
the fundamental representation of $U(N)$.  The $SO(8)$ gauge symmetry
along the infinite worldvolume of the fractional D5 branes is frozen
and recovers the flavor symmetry which rotates the fundamentals of the
effective gauge theory.

\begin{figure}[hbtp]
\begin{center}
\scalebox{0.6}{\input{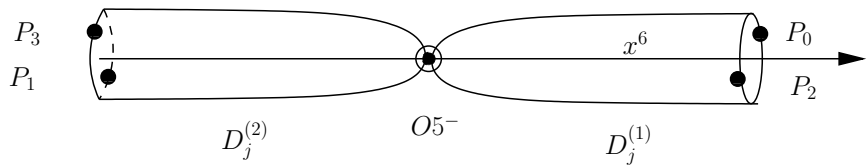}}
\end{center}
\caption{The IIB orientifold action ${\hat \kappa}$ along the locus
  $D_j^{(1)}\cup D_j^{(2)}$. The figure shows an orbit of ${\hat
    \kappa}$ consisting of points $P_i={\hat \kappa}^i(P_0)$, with
  $i=0,1,2,3$ and $P_0$ a point on $D_j^{(1)}$.  }
\label{fiber_action}
\end{figure}

\paragraph{Observation} It is instructive to consider the IIB orientifold action on the
exceptional curves $D_j^{(1)}$ and $D_j^{(2)}$, which are realized in
the plumbing description as $S^1$ fibrations over the half-axes in
$\R^3$ given by $x^4=x^5=0$ and $x^6$ positive or negative. Along
these loci we have $u_1=0$ (for $D_j^{(1)}$) and $x_1=0$ (for
$D_j^{(2)}$) and we can take $x^7=arg(x_1)$ respectively
$x^7=arg(u_1)$.  Then (\ref{U1action}) acts fiberwise by
$x^7\longrightarrow x^7+\theta$, while the orientifold ${\hat \kappa}$
fixes $x^4=x^5=0$ and acts on $x^6$ and $x^7$ as \footnote{For
  convenience, we scale the radius of the $S^1$ fiber to equal one
  (remember that we don't know the metrics anyway).}:
\begin{eqnarray}
x^6&\longrightarrow& -x^6\nn\\
x^7&\longrightarrow& \left[1-{\rm
    sign}(x^6)\right]\frac{\pi}{2}-x^7~~.  \end{eqnarray} The action on $x^7$
corresponds to reflection along one of two orthogonal diameters of the
$S^1$ fiber, depending on whether one acts on a point of $D_j^{(1)}$
or $D_j^{(2)}$ (figure \ref{fiber_action}). The square ${\hat
  \kappa}^2$ acts fiberwise by inversion with respect to the origin of
the $S^1$ fiber (this is the action by half-period shifts on the
periodic coordinate $x^7$): \begin{eqnarray}
\label{shift}
{\hat \kappa}^2:~~(x^4, x^5, x^6, x^7)\longrightarrow (x^4, x^5, x^6,
\pi +x^7)~~.  \end{eqnarray} Notice that the orientifold action does not
preserve the $S^1$ orbits used to perform the T-duality. This makes it
somewhat nontrivial to implement the T-duality explicitly at the level
of conformal field theory (another issue which complicates such an
approach is the degeneration of the $S^1$ fibers, which is responsible
for the appearance of NS5-branes in the IIA description).  It would
be interesting to study this system in more detail through CFT
methods.

\subsection{Geometric description of supersymmetric vacua}

Let us recall the relevant part of the classical moduli space of
supersymmetric vacua, which was discussed in \cite{chiral1}. The vacua
of interest are solutions of the D- and F-flatness constraints, with
the supplementary assumption that the VEV of $\Phi$ is a normal
matrix, i.e. $[\Phi, \Phi^\dagger]=0$.  This condition allows us to
eliminate possible baryonic branches, which are irrelevant for our
purpose.

The condition that $\Phi$ is normal means that it can be diagonalized
through a unitary gauge transformation.  Hence one can take:
\begin{equation}
\label{Phi_diag}
\Phi={\rm diag}(z_1 1_{N_1}\dots z_d 1_{N_d})~~, \end{equation} where $N_j$ are
nonnegative integers such that $\sum_{j=1}^d{N_j}=N$ (we use the
convention that if some $N_j$ vanishes, then the corresponding
eigenvalue does not appear in (\ref{Phi_diag})). Then it was showed in
\cite{chiral1} that the D- and F-flatness constraints imply that the
VEVs of $S, A$ and $Q_f$ must vanish in this family of vacua.  The
unbroken gauge group in such a vacuum is given by
$\prod_{j=1}^d{U(N_j)}$, with the convention that $U(0)$ is the
trivial group.

In the Hanany-Witten construction, such vacua correspond to a
configuration of the type described above, where each ${\cal
  D}^{(\alpha)}_j$ is viewed as a stack of D4-branes with multiplicity
$N_j$. Note that the D4-branes must be located at $x^8+ix^9=z_j$. This
is required in order to preserve ${\cal N}=1$ supersymmetry. In the
T-dual IIB construction, we simply have $N_j$ D5-branes wrapped on
each of the exceptional curves $D_j^{(1)}$ and $D_j^{(2)}$.

\section{The geometric transition and effective superpotential}
\label{gt}

To describe the geometric transition of our system, we follow
\cite{Cachazo_Vafa} by considering the most general log-normalizable
deformation of $X_0$ \begin{equation}
\label{X0gen}
xy=(u-t_0(z))(u-t_1(z))(u-t_2(z))~~ \end{equation} where: \begin{eqnarray}
t_0(z)&=&-\frac{2W'_1(z)+W'_2(z)}{3}\nn\\
t_1(z)&=&\frac{2W'_2(z)+W'_1(z)}{3}\\
t_2(z)&=&\frac{W'_1(z)-W'_2(z)}{3}\nn~~ \end{eqnarray} and $W_1,
  W_2$ are two polynomials (note that $t_0(z)+t_1(z)+t_2(z)=0$).  It
is not hard to check (see Appendix A) that such a deformation preserves
the discrete symmetry (\ref{or0}) if and only if it has the form: \begin{equation}
\label{X}
xy=u(u^2-W'(z)^2+2f_0(z))~~, \end{equation} where $f_0(z)$ is a polynomial of
degree at most $d-1=deg W'-2$. The deformed space has a 3-section
$\Sigma$ (the deformation of $\Sigma_0$) given by $x=y=0$. This has a
rational component $C_1$ with equation $u=0$ and a hyperelliptic piece
$\Sigma^{red}$ given by: \begin{equation}
\label{Sigma_red}
\Sigma^{red}:~~u^2-W'(z)^2+2f_0(z)=0~~.  \end{equation}

After the geometric transition ${\hat X}\longrightarrow
X_0\longrightarrow X$ of \cite{Cachazo_Vafa}, the D5-branes wrapping
the exceptional fibers of ${\hat X}$ are replaced by fluxes through
the $S^3$ cycles created by smoothing, and the orientifold plane is
replaced by the fixed point set $O$ of the action (\ref{or0}) on the
deformed space $X$. This is given by $x=y=u=0$, which is the curve
$C_1$. The D5-branes wrapped over the orientifold fixed locus also
survive the transition.  Thus we end up with a compactification with
NS-NS and R-R fluxes and a $\Z_4$ orientifold which fixes an O5 plane
with worldvolume $\R^{1,3}\times C_1$, together with $8$ fractional
D5-branes wrapping the orientifold fixed `plane'. As above, the RR
charge of the D5-branes cancels the RR charge of the orientifold.  The
worldvolume of these D5-branes carries an $SO(8)$ symmetry, which is
frozen to a global symmetry of the system because the internal part of
the D5-brane worldvolume is non-compact. This gives the geometric
realization of the $SO(8)$ flavor symmetry, which is unbroken after
confinement.

Since the component $C_1:~~u=0$ of the 3-section is unchanged, the
transition can be described as the deformation of the degenerate curve
$\Sigma_0^{red}$ defined by $u^2=W'(z)^2$ to the smooth Riemann
surface (\ref{Sigma_red}). As we shall see below, this `reduced'
geometric process also describes the {\em planar limit} of the
geometric transition associated with the $SO(N)$ theory with symmetric
matter. In fact, the reduced Riemann surface (\ref{Sigma_red})
coincides with the spectral curve governing the strict planar limit of
the matrix model associated to such theories via the Dijkgraaf-Vafa
correspondence \cite{KRS1,Alday,KRS, chiral1}.  Geometrically, the component
$C_1$ of the Riemann surface is a `spectator' during the geometric
transition, with the relevant planar information encoded by the
hyperelliptic curve $\Sigma_{red}$\footnote{In the M-theory lift of
  the IIA Hanany-Witten configuration this shows up in that the middle
  NS-brane remains flat \cite{park3,lll}.}.  This matches the field
theory and matrix model results of \cite{chiral1}. Indeed, it was
shown in \cite{chiral1} that the effective superpotential of our
chiral theory agrees with that of the $SO(N)$ model with symmetric
matter.  The easiest way to see this is to turn on a small positive
Fayet-Iliopoulos parameter in the chiral model, an operation which
cannot affect the effective superpotential since the latter is
protected by holomorphy. Restricting to the vacua of interest (namely
those vacua of the chiral theory for which the matrix $\Phi$ is
normal) one finds \cite{chiral1} that turning on such a
Fayet-Iliopoulos parameter leads to a pattern of gauge symmetry
breaking which recovers the $SO(N)$ theory with symmetric matter as an
effective description of our chiral model (the symmetric field $X$ of
the $SO(N)$ model arises as the symmetric part of $\Phi$, which
remains massless after turning on the Fayet-Iliopoulos parameter).
Then a direct calculation \cite{chiral1} shows that the scales of the
two theories agree, a `miracle' which is due to the particular field
contents under consideration. Hence the effective superpotential of
the chiral theory for this particular set of vacua must agree with
that of the $SO(N)$ model with symmetric matter, a conclusion which
has been verified by computing the two effective superpotentials upon
using the Dijkgraaf-Vafa correspondence \cite{chiral1}. The fact that
the component $C_1$ of our Riemann surface is not affected by the
geometric transition allows us to describe the transition in terms of
the spectral curve $\Sigma^{red}$ of the $SO(N)$ model with symmetric
matter. This is the geometric manifestation of the relation observed
in \cite{chiral1} between the two field theories.

Of course, the common effective superpotential of these theories
receives contributions from both orientable and unorientable planar
diagrams.  In the associated matrix models, the former correspond to
the strict planar limit, while the latter give the subleading
contribution in the $1/{\hat N}$ expansion.  Recall from
\cite{chiral1} that the gaugino superpotential can be expressed as:
\begin{equation}
\label{Weff}
W_{eff}=\sum_{j=1}^d{\left[N_j \frac{\partial F_0}{\partial
      S_j}+4F_1+\alpha_j S_j\right]}~~, \end{equation} where $S_j$ are the
gaugino condensates after confinement in the $U(N_j)$ factors and
$\alpha_j$ are related to the effective gauge couplings in these
factors.  Here $F_0$ and $F_1$ are the leading and subleading
contributions to the free energy of the holomorphic matrix model associated to our chiral
theory: \begin{equation}
F=F_0+\frac{g}{{\hat N}}F_1+O( \left(\frac{g}{\hat N}\right)^2)~~,
\end{equation}
where ${\hat N}$ describes the size of the various matrices involved.
We refer the reader to \cite{chiral1} for the construction and
analysis of this matrix model. As noted in \cite{chiral1}, the
partition function of this model agrees up to a factor with that of
the matrix model associated with the $SO(N)$ theory with symmetric
matter, a fact which can be established by comparing the two
eigenvalue representations. More precisely, one has the relation
\cite{chiral1}: \be F=F_{SO}-\frac{g}{{\hat N}}\ln 2~~.  \ee In the
chiral theory, the effective superpotential arises from planar
diagrams with or without boundaries associated to the fundamentals
$Q_f$.  The former correspond to the closed string contribution to the
free energy after the geometric transition, while the latter arise
from unoriented open strings ending on the eight fractional D5-branes
in the presence of the orientifold. As argued in \cite{vafa}, such
contributions can be computed directly in the twisted model, so they
arise from an orientifolded version of Kodaira-Spencer theory
\cite{BCOV}, which unfortunately does not seem to have been studied.
One could also try to apply the ansatz of \cite{Vafa_or} in order to
extract $W_{eff}$ directly from our geometric realization. However,
remember that the RR charge of the O5-plane is canceled (even locally)
by the charge of the D5-branes wrapping the orientifold fixed locus.
Hence a naive application of the ansatz of \cite{Vafa_or} predicts a
zero contribution to $W_{eff}$ from our (O5, D5-brane) system ! Of
course this cannot be correct, since it contradicts the matrix model
and field theory results of \cite{chiral1}. The problem can be traced
to the fact that the ansatz of \cite{Vafa_or} treats orientifold
planes simply as sources of flux, so it implies that an orientifold
plane can be replaced by a system of D-branes of equal RR charge
density as far as its contribution to the effective superpotential is
concerned. In our model, the D5-branes wrap the orientifold 5-plane
such that the RR charge of the system vanishes locally. However, we
know that this system must bring a nonzero contribution to the
effective superpotential in order to agree with the results of
\cite{or}. Also notice that this is a `degenerate' case in the sense
that our orientifold plane has the same worldvolume as the fractional
D5 branes. It seems that the proposal of \cite{Vafa_or} must be
modified in order to account for our situation. One might need to
relax the assumption that orientifold planes can be treated the same
as D5 branes for the purpose of computing the effective
superpotential.  Such a modified proposal should recover relation
(\ref{Weff}), which was established in \cite{chiral1} through a direct
analysis of generalized Konishi anomalies.

\section{Engineering of the $SO(N)/Sp(N/2)$ theory with \newline
symmetric/antisymmetric matter}
\label{so_eng}

As explained in \cite{chiral1} and recalled above, our theories are
intimately related to the $SO(N)$ theory with symmetric matter. It is
therefore instructive to consider the geometric engineering of such
models within the set-up of \cite{or}. We shall present a geometric
realization of such theories through certain $\Z_2$ orientifolds of
IIB string theory on non-compact Calabi-Yau $A_2$ fibrations in the
presence of D5-branes. More precisely, our IIB background contains a
(disconnected) O5-`plane' sitting in the minimal resolution of such a
fibration, together with D5-branes wrapping the exceptional $\P^1$'s
of the resolution.  By changing the sign of the RR charge of the
orientifold `plane', the set-up discussed below allows us to engineer
both the $SO(N)$ with symmetric matter and $Sp(N/2)$ theories with 
antisymmetric matter.
Thus we shall consider these cases simultaneously. The geometric
engineering discussed below could be of independent interest in light
of the recent analysis of such models within the framework of the
Dijkgraaf-Vafa correspondence \cite{KRS1,Alday,KRS}. We shall also consider the
geometric transition of \cite{vafa, Cachazo_Vafa, Cachazo_Vafa_more}
for our backgrounds. As in \cite{or}, we find that the orientifold
5-`plane' survives the transition (through it becomes connected on the
other `side'), and therefore it contributes to the gaugino
superpotential obtained after confinement in such field theories.

Let us start with the singular $A_1$ fibration $X_{1,0}$ given by:
\begin{equation}
\label{X_10}
X_{1,0}:~~xy=(u-W'(z))(u+W'(z))~~.  \end{equation} This fibration admits the
two-section: \begin{equation}
\label{Sigma_10}
\Sigma_{1,0}:~~x=y=0,~~(u-W'(z))(u+W'(z))=0~~, \end{equation} whose irreducible
components are the rational curves $u=\pm W'(z)$.

The resolution ${\hat X}_1$ can be described globally as the complete
intersection: \begin{eqnarray}
  \beta ( u - W'(z)) &=& \alpha x \nn\\
  \alpha (u +W'(z)) &=& \beta y \\
  (u-W'(z))(u+W'(z)) &=& xy~~\nn \end{eqnarray} in the ambient space
$\P^1[\alpha,\beta]\times \C^4[z, u, x, y]$.  The exceptional $\P^1$'s
sit above the singular points of $X_{1,0}$, which are determined by
$x=y=u=0$ and $z=z_j$, where $z_j$ are the roots of $W'$. We let $D_j$
denote the exceptional $\P^1$ sitting above $z_j$.  The resolved space
admits the $U(1)$ action: \begin{equation}
\label{U1red}
([\alpha, \beta], z, u, x, y) \longrightarrow ([e^{-i\theta} \alpha,
\beta], z, u,e^{i\theta} x, e^{-i\theta}y)~~, \end{equation} which will be of
interest below.,

We next add an orientifold. Consider the holomorphic $\Z_2$ action:
\begin{equation}
\label{red_or}
{\hat k}_1:~~([\alpha,\beta],z,u,x,y) \longrightarrow ([-\beta,
\alpha], z, -u, y, x)~~, \end{equation} which is a symmetry of ${\hat X}_1$
stabilizing each exceptional curve $D_j$. It projects to the following
involution of $X_{1,0}$: \begin{equation}
\label{red_or_proj}
\kappa_1:~~(z,u,x,y) \longrightarrow (z, -u, y,x)~~, \end{equation} whose fixed
point set is given by: \begin{equation}
\label{Olocus}
O_{1,0}: x=y,~~u=0,~~x^2+W'(z)^2=0~~.  \end{equation} The locus $O_{1,0}$ is a
reducible curve with rational components $x=\pm i W'(z)$, which touch
each other at the points $(x,z)=(0,z_j)$.  The fixed point set ${\hat
  O}_1$ of (\ref{red_or}) is the disjoint union of the proper
transforms of these curves: \begin{equation} {\hat
  O}_1:~~x-y=u=x^2+W'(z)^2=0~~,~~\frac{\alpha}{\beta}=\pm i~~.  \end{equation}
Thus the orientifold action on the resolved space determines a
(disconnected) orientifold 5-`plane'.

It is not hard to check that this construction engineers the $SO(N)$
or $Sp(N/2)$ theory with symmetric or antisymmetric matter, respectively.  
The two cases arise by
taking the O5 `plane' to have positive/negative RR charge.  The field
content can be extracted geometrically or by a fractional brane
construction. Alternatively, one can use T-duality to map our IIB
background to one of the known Hanany-Witten realizations of these
field theories \cite{Csaki}, and we shall use this approach below.  This is
essentially an application of \cite{OT}, supplemented with the ansatz
proposed in \cite{or} for the T-dual coordinates.

\subsection{The T-dual configuration}

To extract the T-dual Hanany-Witten configuration, we use a local
description valid on a subset $\tilde X_1 \subset \hat X_1$.  In the
present case, it is given by two copies $U_0$ and $U_1$ of $\C^3$ with
coordinates $(x_i, u_i, z_i)$ ($i=0,1$) which are glued together
according to: \begin{equation} (x_1, u_1, z_1) = (\frac{1}{u_0}, x_0
  u_0^2 -2 W'(z_0) u_0, z_0) \end{equation} The resolution map $\tau$
is given by:
\begin{eqnarray}
(z,u,x,y) =& ( z_0, x_0 u_0 - W'(z_0), x_0 , u_0(x_0 u_0 - 2 W'(z_0)) ) \,,\\
          =& ( z_1, x_1 u_1 + W'(z_1), x_1(x_1 u_1 + 2 W'(z_1)), u_1 )\,.
\end{eqnarray}
The $U(1)$ action (\ref{U1red}) takes the form: \begin{equation}
\label{U1red_local}
(z_i, u_i, x_i ) \longrightarrow (z_i, e^{-i\theta}u_i,
e^{i\theta}x_i)~~, \end{equation} with fixed point set given by the union of
rational curves $x_0=u_0=0$ and $x_1=u_1=0$.  It stabilizes the
exceptional curves $D_j:~x_0=u_1=z-z_j=0$ of the resolved fibration.

The Hanany-Witten construction is obtained by T-duality with respect
to the circle orbits of (\ref{U1red_local}). Following \cite{or}, we
shall use the following ansatz for the T-dual coordinates:
\begin{equation} x^4 + i x^5 = x_0 u_0 - W'(z_0)=x_1 u_1 + W'(z_1)~~,
  ~~x^6=\frac{1}{2}(|x_1|^2-|u_0|^2)~~,~~z=x^8+ix^9~~~~~~~~~~~~~~~
\label{base_coords_a1}
\end{equation} together with the periodic coordinate $x^7$ along the orbits of
the $U(1)$ action (\ref{U1red_local}).

Expressing the fixed point set of (\ref{U1red_local}) in the
coordinates (\ref{base_coords_a1}), we find that the T-dual background
contains two NS5-branes ${\cal N}_0$ and ${\cal N}_1$ sitting at:
\begin{eqnarray} {\cal N}_0:~~x^4+ix^5=-W'(z)~~,~~x^6=+\infty
\end{eqnarray} and: \begin{equation} {\cal
    N}_1:~~x^4+ix^5=+W'(z)~~,~~x^6=-\infty~~, \end{equation} as well
as D4-branes ${\cal D}_j$ stretching between the NS5-branes at $z=z_j$.

\begin{figure}[hbtp]
\begin{center}
  \scalebox{0.7}{\input{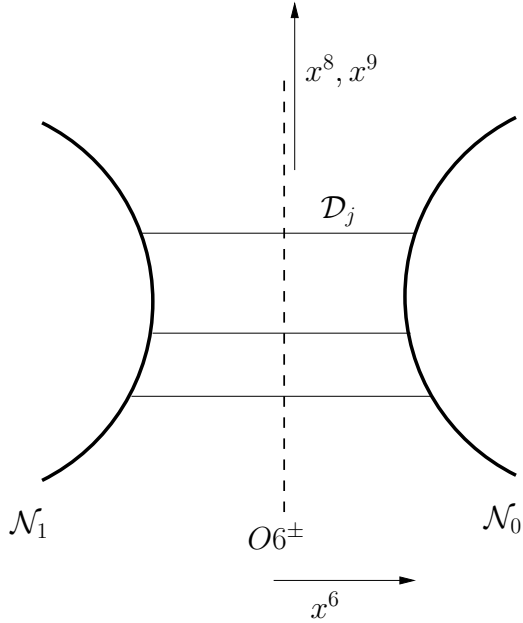}}
\end{center}
\caption{Brane configuration for the $SO(N)/Sp(N/2)$ theories with 
symmetric/antisymmetric matter.
The outer NS5-branes are
  bent in the directions $x^4$ and $x^5$, which cannot be shown
  properly in this two-dimensional figure. The $SO(N)/Sp(N/2)$ gauge
  groups correspond to positive/negative charge of the orientifold
  6-plane.}
\label{O6plane}
\end{figure}

The orientifold (\ref{red_or}) acts in local coordinates as:
\begin{equation}
\label{red_local}
(z_0, x_0 , u_0) \longleftrightarrow (z_1, u_1, -x_1)~~.  \end{equation} The
fixed point set is $u_0^2+1=x_0+W'(z)u_0=0$, which is the union of two
disjoint rational curves mentioned above.  Using
(\ref{base_coords_a1}) we find that under T-duality this locus maps to
an O6-plane sitting at $x^4=x^5=x^6=0$ (figure \ref{O6plane}). Notice
the simple form of the dual O6-plane, despite the fact that the
original O5-plane in the resolved space ${\hat X}_1$ has two
disconnected components.  This is due to nonlinearity of
the map (\ref{base_coords_a1}).

The brane configuration of figure \ref{O6plane} is one of the
Hanany-Witten systems realizing the SO/Sp theories with one 
symmetric/antisymmetric
chiral multiplet. This establishes the fact that the IIB background
considered above engineers these theories.

\subsection{Description after the geometric transition}

After the geometric transition of \cite{vafa, Cachazo_Vafa,
  Cachazo_Vafa_more}, the Calabi-Yau space (\ref{X_10}) is deformed
to: \begin{equation}
\label{X1}
X_{1,0}:~~xy=u^2-W'(z)^2+2f_0(z)~~, \end{equation} where $f_0(z)$ is a polynomial
of degree at most $d-1$.  This fibration admits the two-section: \begin{equation}
\label{Sigma_1}
\Sigma_{1}:~~x=y=0,~~u^2-W'(z)^2+2f_0(z)=0~~, \end{equation} which coincides with
the reduced component (\ref{Sigma_red}) obtained after transition in
the chiral theory.  The D5-branes wrapping the exceptional divisors of
the resolution are replaced by fluxes, but the deformed space
(\ref{X1}) is still invariant under the $\Z_2$ action
(\ref{red_or_proj}) so the orientifold 5-plane survives the
transition. Its internal part is deformed to the irreducible curve:
\begin{equation} O_1:~~x=y~~,~~x^2+W'(z)^2-2f_0(z)=0~~.  \end{equation}

The Riemann surface (\ref{Sigma_1}) arises naturally in the confining
phase of the \linebreak $SO(N)/Sp(N/2)$ theory with symmetric/antisymmetric 
matter
\cite{KRS1,Alday,KRS, chiral1}.  As explained in \cite{Alday,KRS}, this 
curve can be
extracted by analyzing the generalized Konishi anomalies of such
theories, and coincides with the spectral curve which governs the
large N limit of the matrix model associated to them via the
Dijkgraaf-Vafa correspondence. As in \cite{or}, the orientifold
5-`plane' in our geometric engineering survives the transition and
brings a nontrivial contribution to the effective superpotential.

\section{Conclusions}
\label{cl}

We discussed the geometric engineering and large N transition for the
chiral $\cn=1$ field theory containing one adjoint, one antisymmetric,
one conjugate symmetric and eight fundamental chiral multiplets.  This
turned out to be considerably more complex then the much better
studied case of non-chiral models. Beyond D5-branes wrapping the
exceptional curves of a resolved $A_2$ fibration, our set-up involves
the introduction of a $\Z_4$ orientifold together with eight
fractional D5-branes required by cancellation of RR tadpoles. This
corresponds to cancellation of the chiral anomaly by the eight
fundamentals of the field theory model.

Although it is straightforward to perform the large N transition for
such backgrounds, it turns out that the $\Z_4$ orientifold and
fractional D5 branes survive the transition, and it is far from
obvious how to compute their contributions to the effective
superpotential by string theory methods. This seems to require a
nontrivial modification of the ansatz of \cite{Vafa_or}, or perhaps a
more systematic approach through Kodaira-Spencer theory \cite{BCOV}.  At the
moment we have to rely on the field theoretic arguments of
\cite{chiral1}, which determine the effective superpotential through a
direct analysis of generalized Konishi anomalies and express it in
terms of a holomorphic matrix model as expected from the
Dijkgraaf-Vafa correspondence.

There are a couple of aspects of our construction which deserve
further study.  In section 3 we used local RR charge conservation to
argue that adding eight fractional D5-branes in our IIB background
amounts to canceling RR tadpoles.  It would be interesting to check
this directly by an explicit study of the tadpole cancellation
constraints. Since this is quite nontrivial in our curved background,
one could follow the approach of \cite{Uranga_probes} which relies on
testing consistency of the worldvolume theory of various D-brane
probes.

One subtle aspect found in \cite{chiral1} is a mismatch between the
number of flavors in the field theory and associated holomorphic
matrix model.  As discussed there, the associated matrix model is
consistent only if the number of matrix model flavors equals two.  It
would be interesting to give a direct derivation of the matrix model
of \cite{chiral1} from the B-twisted model associated to our
backgrounds.

\acknowledgments{ K. L. would like thank G. Honecker and A. Uranga for
  discussions.  This work was supported by DFG grant KL1070/2-1.  R.
  T. is supported by DOE Contract DE-AC03-76SF00098 and NSF grant
  PHY-0098840.}

\appendix

\section{Relation to general log-normalizable Calabi-Yau $A_2$ fibrations}
Consider a singular $A_2$ fibration of the form (\ref{X0gen}).  The
symmetry (\ref{or0}) is preserved provided that: \begin{equation}
\label{tprod}
\prod_{j=0}^2(u-t_j(z))=\prod_{j=0}^2(u+t_j(z))\Longleftrightarrow
t_0t_1t_2\equiv 0 \end{equation} This leads to the possibilities: \begin{eqnarray}
\label{cases}
t_0\equiv 0&\Longleftrightarrow&
W'_1(z)=-\frac{1}{2}W'_2(z):=W'(z)\Longrightarrow
t_2(z)=-t_1(z)=W'(z)\nn\\
t_1\equiv 0&\Longleftrightarrow&
W'_2(z)=-\frac{1}{2}W'_1(z):=W'(z)\Longrightarrow t_2(z)=-t_0(z)=-W'(z)\\
t_2\equiv 0&\Longleftrightarrow& W'_1(z)=W'_2(z):=W'(z)\Longrightarrow
t_1(z)=-t_0(z)=W'(z)\nn~~.  \end{eqnarray} In each of these cases, we have
$t_0t_1+t_1t_2+t_2t_0=-W'(z)^2$ and the hypersurface (\ref{X0gen})
reduces to the form (\ref{X0}).

Consider now a general log-normalizable deformation of (\ref{X0gen}):
\begin{equation}
\label{Xgen}
xy=u^3-p(z)u-q(z) \end{equation} where
$p(z)=t_0(z)^2+t_1(z)^2+t_0(z)t_1(z)-\psi_0(z)-\psi_1(z)$ and
$q(z)=-t_0(z)t_1(z)(t_0(z)+t_1(z))+t_0(z)\psi_1(z)+
t_1(z)\psi_0(z)-g(z)$, where $\psi_0,\psi_1$ and $g$ are polynomials
of degrees $d-1$. Invariance of (\ref{Xgen}) under (\ref{or0})
requires $q\equiv 0$, which together with (\ref{tprod}) implies: \begin{equation}
\label{ppart}
p(z)=W'(z)^2+f(z) \end{equation} where $f(z):=-\psi_0(z)-\psi_1(z)$ is an
arbitrary polynomial of degree $d-1$. Relation (\ref{ppart}) holds in
each of the cases (\ref{cases}).  Thus equation (\ref{Xgen}) reduces
to the form (\ref{X}).

\section{Fractional brane construction}
\label{fractional}

For constant $W$, our singular Calabi-Yau threefold becomes the
trivial $A_2$ fibration: \begin{equation}
\label{X00}
xy=u^3~~,~~z={\rm arbitrary} \end{equation} which is simply the product space
$\C\times (\C^2/\Z_3)$.  We can view this as the orbifold
$\C^3/\Z_3$ with action: \begin{equation}
\label{rho_orbf}
\rho(\theta):=(z_0, z_1,z_2) \rightarrow (z_0, \theta z_1, \theta^{-1}
z_2) \end{equation} where $\theta=e^{2\pi i/3}$ and $z_0$, $z_1$, $z_2$ are the
affine coordinates of $\C^3$.  Introducing the invariant coordinates:
\begin{equation}
\label{ic}
x=z_1^3~~,~~y=z_2^3~~,~~u=z_1z_2~~,~~z:=z_0 \end{equation} the factor $\C^2/\Z_3$
takes the standard form: \begin{equation} xy=u^3~~.  \end{equation} and we recover the
description (\ref{X00}).

To identify the physics in this limit, let us set $W=\lambda W_0$ and
take $\lambda \rightarrow 0$.  When $\lambda=0$, the superpotential
has the entire $z$-plane as a critical set, and (\ref{X00}) can be
resolved by double blow up of every fiber. In particular, there is no
distinguished point in the base specifying the location of the two
D5-branes wrapped over the exceptional $\P^1$'s --- we now have two
such exceptional curves above every point in the base.  Physically,
$W$ provides a potential barrier between the states described by given
configurations of wrapped D5-branes. In the adiabatic (WKB) limit
$\lambda\rightarrow 0$, this barrier becomes vanishingly small and
tunneling allows the D5-branes to spread out in the $z$ direction.
Accordingly, blowing down all fibers after setting $\lambda=0$ will
lead to fractional D3-branes which are delocalized in the $z$
direction, thereby spanning a worldvolume equal to $\R^{1,3}\times
\C$. On the other hand, the D5-branes wrapping the rational curve
${\hat C}_1$ survive trivially in this limit, yielding fractional
D5-branes with worldvolume $\R^{1,3}\times \C$.

\begin{figure}[hbtp]
\begin{center}
  \scalebox{0.6}{\input{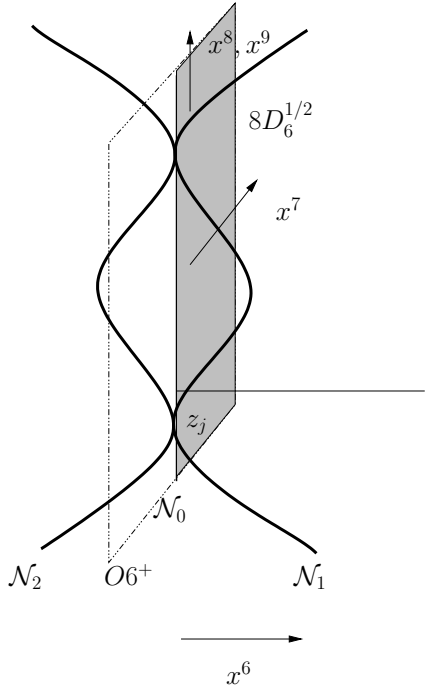}}
\end{center}
\caption{Singular limit in the Hanany-Witten picture.}
\label{lim_config}
\end{figure}

It is instructive to consider this limit in the T-dual IIA
description.  Remember that ${\cal N}=1$ supersymmetry requires that
the D4-branes of the Hanany-Witten configuration are localized at the
critical points $z_j$ of $W$ in the direction $z=x^8+ix^9$ (see figure
\ref{chiral_config}).  Since the length of the D4-branes in the
direction $x^6$ is related to the size of the exceptional $\P^1$'s of
the T-dual model, the singular limit of our $A_2$ fibration
corresponds to the outer NS5-branes touching each other at the points
$z=z_j$ \footnote{This is not visible in the coordinates used in
  Subsection 3.2, which are chosen such that the outer NS branes sit
  at infinity. However, one can redefine these coordinates to place
  the outer NS5 branes at finite distance.} (figure \ref{lim_config}).
In this limit, all D4-branes become tensionless and can be viewed as
codimension two massless excitations inside the five-dimensional
boundary of the half D6-branes. Note that the gauge coupling
$g_{YM}^2=g_s$ on the half D6-branes becomes infinite along this
boundary, since there is an NS5-brane sitting there. Hence all
solitonic excitations of the worldvolume theory of the D6 branes
become massless along this boundary, and we can identify the massless
D4 branes with solitonic objects confined to the boundary. A
nontrivial $W$ provides a potential barrier for the movement of such
objects in the $z$ direction, which becomes vanishingly small in the
limit $W'\equiv 0$. In this limit, the tensionless D4 branes become
delocalized along the boundary of the half D6 branes.

Returning to the IIB picture, consider the field theory along the
common part $\R^{1,3}$ of the branes' worldvolume. Since we are mostly
interested in the massless field content, it suffices to look at open
string modes associated with the coordinates $x^0\dots x^3$, which are
insensitive to the fact that the fractional D3 branes are delocalized
in the direction $z$. Thus we can use the standard methods of
\cite{Douglas_moore} in order to extract the massless field content.
For this, let us consider the Chan-Paton representation
$R=\oplus_{i=1}^2{E_i\otimes R_i}$, where $E_i$ are some
finite-dimensional complex vector spaces and $R_i$ are the irreducible
representations of $\Z_3$, i.e. copies of $\C$ carrying the actions:
\begin{equation} {\hat R}_i(\theta)=\theta^i~~.  \end{equation}

The superfield potential ${\bf V}$ and chiral superfields $Z_0\dots
Z_2$ of the worldvolume theory transform in the $\Z_3$ representations
$End(R)$ and $\rho \otimes End(R)$ respectively (here $\rho$ is the
geometric representation given in (\ref{rho_orbf})).  This gives the
orbifold projection: \begin{eqnarray}
  {\bf V}_i\,^j &=& \theta^{i-j} {\bf V}_i\,^j\,\\
  (Z_1)_i\,^j &=& \theta^{i-j+1} (Z_1)_i\,^j\,\\
  (Z_2)_i\,^j &=& \theta^{i-j-1} (Z_2)_i\,^j~~ \end{eqnarray} where
${\bf V}_i\,^j$ and $(Z_k)_i\,^j$ are the components along
$Hom(E_j,E_i)$.  The surviving components are ${ \bf V}_i\,^i$,
$(Z_0)_i,^i$ and $(Z_1)_1\,^2$, $(Z_2)_2\,^1$.

We next consider the branes wrapping ${\hat C}_1$. Since these are
T-dual to the half D6-branes of the Hanany-Witten construction, they
should transform nontrivially under the action of the square root
$\xi:=\theta^{1/2}= e^{i\pi/3}$, which generates a $\Z_6$ group. We
implement this by considering the Chan-Paton representation
$\Gamma:=F\otimes \Gamma_3$, where $F$ is some finite-dimensional
complex vector space and: \begin{equation} {\hat \Gamma}_i(\xi)=\xi^i
\end{equation} give the the irreps $\Gamma_i$ of $\Z_6$. In terms of
the $\Z_6$ generator, the action (\ref{rho_orbf}) takes the form:
\begin{equation}
\label{rho_xi}
{\tilde \rho}(\xi):(z_0,z_1,z_2)\rightarrow (z_0, \xi^2 z_1, \xi^4
z_2)~~.  \end{equation} This is a non-effective $\Z_6$ action on $\C^3$, with a
trivially acting $\Z_2$ subgroup generated by $\xi^3=-1$. The original
D5-branes transform in the $\Z_6$ representations $R_1=\Gamma_2$ and
$R_2=\Gamma_4$, while $\Z_i$ have the geometric transformations
(\ref{rho_xi}).

\begin{figure}[hbtp]
\begin{center}
  \scalebox{0.5}{\input{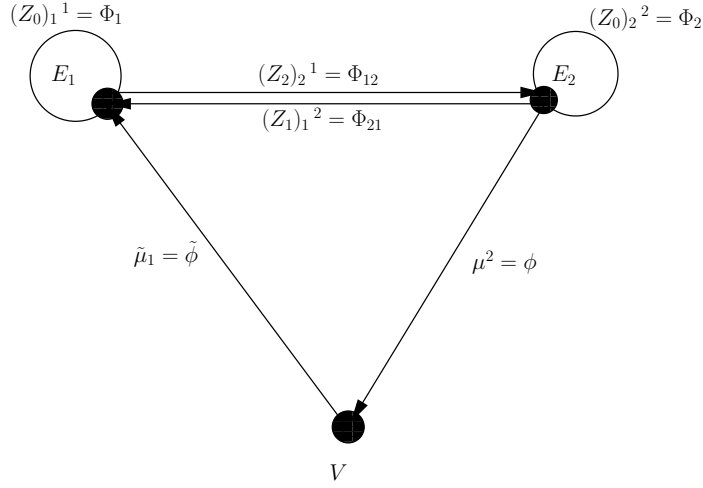}}
\end{center}
\caption{The quiver representation before orientifolding. Here $\Phi_{ij}$
etc are the notation used in Appendix \ref{eng_no_pot}.}
\label{quiver}
\end{figure}

After introducing the new branes, we have chiral superfields $\mu,
{\tilde \mu}$ with Chan-Paton representations $Hom(R, \Gamma)$ and
$Hom(\Gamma,R)$, as well as fields in the representation
$End(\Gamma)$. The latter will be unimportant for our purpose since
they live on the central D5-brane, whose worldvolume in the internal
directions remains infinite even after turning on the superpotential
$W$ and resolving the singular fibers.  This means that such fields
will be frozen to the values specified by their equations of motion.
For the fields $\mu, {\tilde \mu}$ we shall consider the geometric
transformations: \begin{equation} (\mu,{\tilde \mu})\rightarrow
  (\xi^{-1} \mu, \xi^{-1} {\tilde \mu})~~ \end{equation} under the
action (\ref{rho_xi}). This gives the $\Z_6$ projections:
\begin{eqnarray}
  \mu&=& \xi^{-1} {\hat \Gamma}(\xi) \mu {\hat R}(\xi^2)^{-1}\nn\\
  {\tilde \mu} &=& \xi^{-1} {\hat R}(\xi^2) {\tilde \mu} {\hat
    \Gamma}(\xi)^{-1}~~, \end{eqnarray} with the solution:
\begin{eqnarray} \mu=\left[\ba{cc}0 & \mu^1\ea\right]~~,~~{\tilde
    \mu}=\left[\ba{c}{\tilde \mu}_2\\0\ea\right]~~, \end{eqnarray}
where $\mu^1\in Hom(E_1,F)$ and ${\tilde \mu}_2\in Hom(F,E_2)$. The
surviving fields form the quiver representation shown in figure
\ref{quiver}. This recovers the desired field content in the absence
of the orientifold projection, in agreement with the work of
\cite{park1}.

We next consider a $\Z_4$ orientifold of this configuration, whose
geometric action is given by: \begin{equation}
\label{or_orbf}
(z_0, z_1, z_2) \rightarrow (z_0, -z_2, z_1)~~.  \end{equation} In the invariant
coordinates (\ref{ic}), this becomes $(z, u, x, y)\rightarrow (z, -u,
-y, z)$, which agrees with the action used above for geometric
engineering. This orientifold makes sense only if $\dim E_1=\dim
E_2:=N$, which assume from now on.

\begin{figure}[hbtp]
\begin{center}
  \scalebox{0.5}{\input{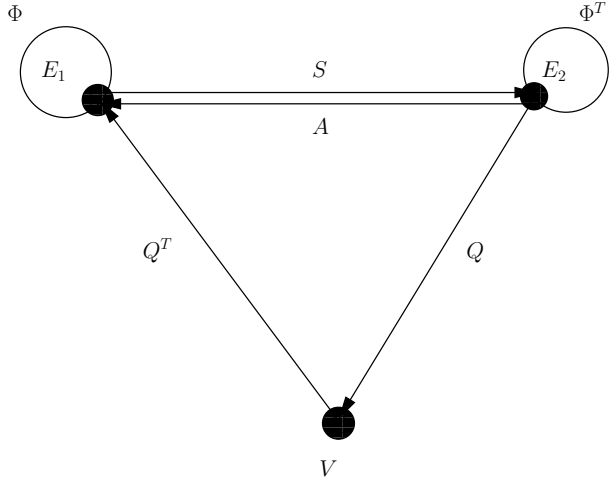}}
\end{center}
\caption{Projected quiver representation. }
\label{quiver_proj}
\end{figure}

For the Chan-Paton actions, we take: \begin{eqnarray}
  \gamma=\left[\ba{cc} 0 & {\bf 1}_N\\{\bf 1}_N &0 \ea\right]~~,
  ~~\eta= 1_{N_F}~~, \end{eqnarray} where $N_F:=\dim F$. The
orientifold projections are: \begin{eqnarray}
  {\bf V} &=& -\gamma {\bf V}^T \gamma^{-1}\\
  Z_0 &=& \gamma Z_0^T \gamma^{-1}\\
  Z_1 &=& -\gamma Z_1^T \gamma^{-1}\\
  Z_2 &=& \gamma Z_2^T\gamma^{-1}\\
  {\tilde \mu}&=& \gamma \mu^T \eta^{-1}~~.  \end{eqnarray} This
orientifold action exchanges nodes $E_1$ and $E_2$ of the quiver,
while fixing the node labeled $F$.  Writing it in components, we find
(figure \ref{quiver_proj}): \begin{eqnarray}
\label{frac_proj}
&& {\bf V}_2\,^2=-[{\bf V}_1\,^1]^T:={\bf V}~~,~~(Z_0)_2\,^2=[(Z_0)_1\,^1]^T:=\Phi\nn\\
&& (Z_1)_1\,^2=-[(Z_1)_1\,^2]^T:=A~~,~~(Z_2)_2\,^1=[(Z_2)_2\,^1]^T:=S~~\\
&& {\tilde \mu}_2=-(\mu^1)^T:=Q\nn~~.  \end{eqnarray}

One can also consider the vector fields $\nu \in End(\Gamma,\Gamma)$
on the central brane, with the trivial orbifold projection and the
orientifold projection $\nu=-\eta \nu^T\eta^{-1}$. This shows that the
gauge group on the worldvolume of the central brane is projected to
$SO(N_F)$.  Since this brane has infinite volume even after resolving
the singularities and turning on $W$, its gauge symmetry is frozen to
a global $SO(N_F)$ invariance of the effective action of our system.

The first equation in (\ref{frac_proj}) implies that the gauge
transformations $U_1$ and $U_2$ are identified as $U_2=U_1^{-T}:=U$.
This produces the desired $U(N)$ gauge symmetry.  We are left with the
antisymmetric and symmetric fields $A$, $S$, the adjoint field $\Phi$
and the $N_F$ fundamental fields given by the components of $Q\in
Hom(W,E_1)$ along a basis of $W$.  Because the gauge group elements
associated to $E_0$ and $E_1$ are identified as $U_2=U_1^{-T}:=U$, the
fields $A$ and $S$ transform in the antisymmetric and conjugate
symmetric representations of $U(N)$.  It is easy to check that the
only cubic terms in the superpotential which involve $Q, S$ or $A$ and
are consistent with this field content and gauge invariance are of the
form $\tr (Q^T S Q)$ and $\tr (S\Phi A)$. Of course, RR tadpole
cancellation requires $N_F=8$, as argued independently in section 3.

We also note that an antisymmetric choice 
\begin{equation}
  \label{eq:antisymgamma}
 \gamma=\left[\ba{cc} 0 & {\bf 1}_N\\-{\bf 1}_N &0 \ea\right]~ 
\end{equation}
is physically equivalent. The differences are that with this choice
the gauge transformations are identified according to $U_1=U_2^T$ and that
$Z_2$ projects onto the symmetric field $S$ whereas $Z_1$ 
projects onto the antisymmetric field $A$. In the T-dual Hanany-Witten
construction this corresponds to the choice of having positive O6-plane
charge along the positive or negative $x^7$ direction.

\section{Engineering without a tree level potential}
\label{eng_no_pot}

Let us now sketch the geometric interpretation of the fractional brane
construction.  Consider the toric resolution $(\C^4-{\cal
  Z})/(\C^*)^2$ of the $A_2$ singularity $\C^2/\Z_3$, with charge
matrix:
\begin{equation}
\label{charge}
Q=\left[\begin{array}{cccc}1&-2&1&0\\0&1&-2&1\end{array}\right]~~ \end{equation}
and toric generators given by the columns of: \begin{equation}
G=\left[\begin{array}{cccc}1&1&1&1\\0&1&2&3\end{array}\right]~~.  \end{equation}
The generators correspond to homogeneous coordinates which we denote
by $w_1\dots w_4$.  The exceptional set is ${\cal Z}=\{w_1=w_3=0\}\cup
\{w_2=w_4=0\}\cup \{w_1=w_4=0\}$.  We let $D_j=(w_j)$ be the toric
divisors.  Then $D_2=D^{(1)}$ and $D_3=D^{(2)}$ are the exceptional
$\P^1$'s, while $D_1$ and $D_4$ are non-compact.

In the symplectic quotient description, we have the reduction of
$\C^4$ with respect to the $U(1)^2$ action defined by (\ref{charge})
with moment map equations: \begin{eqnarray}
  |w_1|^2-2|w_2|^2+|w_3|^2&=&\zeta_1\nn\\
  |w_2|^2-2|w_3|^2+|w_4|^2&=&\zeta_2~~, \end{eqnarray} where $\zeta_j$
are some positive levels. Setting $\zeta_1=\zeta_2=0$ recovers the
$A_2$ singularity, which is described by the invariants:
\begin{eqnarray}
  x&=&w_1^3 w_2^2w_3\nn\\
  y&=&w_2w_3^2w_4^3\\
  u&=&w_1w_2w_3w_4~~,\nn \end{eqnarray} subject to the relation
$xy=u^3$.

The orientifold action on the minimal resolution takes the form
\footnote{In the symplectic quotient description, this is a symmetry
  if we set $\zeta_1=\zeta_2=\zeta$.} : \begin{equation}
\label{Yaction}
(w_1,w_2,w_3,w_4)\longrightarrow (w_1,w_3,-w_2,w_4) \end{equation} The only fixed
fixed point is $p:=[1,0,0,1]$, which is the intersection of $D_2$ and
$D_3$. We have the intersections: \begin{equation} D_1\cap
D_2=\{q\}=\{[0,0,1,1]\}~~,~~D_4\cap D_3=\{r\}=\{[1,1,0,0]\}~~.  \end{equation} As
expected, the action (\ref{Yaction}) permutes the compact divisors.

Consider two D5-branes along $\R^{1,3}\times D_2$ and $\R^{1,3}\times
D_3$ with trivial Chan-Paton bundles ${\cal E}_1$ and ${\cal E}_2$
such that ${\hat \kappa}^*({\cal E}_1)\approx {\cal E}_2$ and ${\hat
  \kappa}^*({\cal E}_2)\approx {\cal E}_1$.  Then the orbifold action
${\hat \kappa}^2$ maps ${\cal E}_j$ into itself in the sense that
$({\hat \kappa}^*)^2({\cal E}_j)\approx {\cal E}_j$.  We also consider
a D5-brane with worldvolume $\R^{1,5}\times \{p\}$ , which carries a
Chan-Paton bundle ${\cal F}$. The massless states of the target space
theory arise along the locus $\R^{1,3}\times \{p\}$, which is the
intersection of all worldvolumes. Therefore, it suffices to
concentrate on the fibers $E_j$, $F$ of ${\cal E}_j$ and ${\cal F}$ at
the point $p$.  The massless fields are associated with morphisms
$\Phi_i\in End(E_i)$ and $\Phi_{12}\in Hom(E_1, E_2)$, $\Phi_{21}\in
Hom(E_2, E_1)$ as well as $\phi_1 \in Hom(E_1, F)$, $\phi_2 \in
Hom(E_2, F)$ and ${\tilde \phi_1}\in Hom(F, E_1)$, $\tilde \phi_2 \in
Hom(F,E_2)$.

For simplicity, we shall endow the vector spaces $E_j$ and $V$ with
Hermitian metrics (this is not strictly necessary, but it allows us to
formulate the orientifold projection in a more traditional manner).
Then the orientifold action is implemented by invertible maps:
\begin{eqnarray}
\label{linearizations}
&& \gamma_0:F\rightarrow F\\
&& \gamma_1:E_1\rightarrow E_2\nn\\
&& \gamma_2:E_2\rightarrow E_1~~.
\end{eqnarray}

Let us first consider the wrapped D5 branes. In this sector we have
$\gamma_1\gamma_2^{-T}={\bf 1}_{E_1}$\footnote{Using 
$\gamma_1\gamma_2^{-T}=-{\bf 1}_{E_1}$ leads to 
equivalent results.}.  Using $\gamma_1$ to identify
$E_1$ and $E_2$ to a vector space we call $E$, we can set
$\gamma_1=\gamma_2={\bf 1}_{E}$.  Then the orientifold projection
takes the form (\ref{frac_proj}):
\begin{eqnarray}
  && \Phi_2=\gamma_{1} \Phi_1^T \gamma_{1}^{-1}=\Phi_1^T:=\Phi\\
  && \Phi_{21}=\gamma_{2}\Phi_{21}^T\gamma_{1}^{-1}=\Phi_{21}^T:=S~~\\
  &&\Phi_{12}=-\gamma_1\Phi_{12}^T\gamma_{2}^{-1}= -\Phi_{12}^T:=A~~.
\end{eqnarray}
The orientifold projection on the vector fields ${\bf V}_1$ and ${\bf
  V}_2$ on the two wrapped D5 branes is: \be {\bf V}_2=-\gamma_{1}
{\bf V}_1^T \gamma_{1}^{-1}=-{\bf V}_1^T:={\bf V}~~.  \ee This shows
that the $U(N)$ gauge transformations on the wrapped D5 branes are
identified as: \be U_2=U_1^{-T}:=U~~, \ee thereby giving a $U(N)$
gauge group under which $A$ and $S$ transform in the antisymmetric and
conjugate symmetric representations.

We next consider the orientifold projection on the D5-branes carrying
the trivial Chan-Paton bundle ${\cal F}=F\times {\hat C}_1$.  Since
${\hat C}_1$ is invariant under the $\Z_2$ orbifold, the space $F$
carries a representation of $\Z_2$ and thus it can be decomposed as
$F=F_+\oplus F_-$, where the $\Z_2$ generator acts as $\pm {\bf 1}$ on
the subspaces $F_\pm$.  Because the orientifold generator squares to
the $\Z_2$ orbifold generator, the invertible map $\gamma_0$ must
fulfill:
\begin{equation} \gamma_0 \gamma_0^{-T} = \left(
  \begin{array}{cc}
{\bf 1}_{F_+}&0\\
0&-{\bf 1}_{F_-}
  \end{array}
\right)~~.  \end{equation}
One can choose:
\begin{equation}
\label{gamma0}
\gamma_0 = \left(
  \begin{array}{cc}
\gamma_{0,s}&0\\
0&\gamma_{0,a}
  \end{array}\right)
\end{equation} with $\gamma_{0,s}\gamma_{0,s}^{-T} ={\bf 1}_{F_+}$ and
$\gamma_{0,a}\gamma_{0,a}^{-T} =-{\bf 1}_{F_-}$.  Considering the
vector field ${\bf V}$ living on the fractional D5-branes, the $\Z_2$
orbifold projection implies that only the diagonal sectors ${\bf
  V}^+={\bf V}^{++}$ and ${\bf V}^-={\bf V}^{--}$ survive.  Then the
orientifold projection gives:
\begin{eqnarray}
  {\bf V}^{+} &=& - \gamma_{0,s} ({\bf V}^{+})^T \gamma_{0,s}^{-1}\\
  {\bf V}^{-} &=& - \gamma_{0,a} ({\bf V}^{-})^T \gamma_{0,a}^{-1}~~.
\end{eqnarray}
The result is an $SO$ gauge group in the $+$ sector and an $Sp$ gauge
symmetry in the $-$ sector. Thus we obtain two types of fractional D5
branes, which carry $SO$ respectively $Sp$ gauge symmetry on their
worldvolume.  To cancel the RR tadpoles, we need eight more fractional
D5-branes in the sector $+$ than in the sector $-$.  In the
Hanany-Witten picture, the choice (\ref{gamma0}) corresponds to a more
general configuration which contains half D6 branes stretching in the
positive $x_7$ direction as well as half D6 branes stretching in the
negative $x_7$ direction. The first are T-dual to fractional D5 branes
in the sector $+$, while the second correspond to fractional D5-branes
in the sector $-$.  Notice that one can form bound states of pairs of
fractional branes belonging to the two sectors, giving whole D5-branes
which can move away from the $\Z_2$ orbifold fixed locus ${\hat C}_1$.
In the Hanany-Witten construction, this corresponds to the
recombination of a half D6 brane stretching in the positive $x_7$
direction with a half D6 brane stretching in the negative $x_7$
direction to a whole D6-brane which can move away from the central
NS5-brane.  In the case of interest for us, one has eight fractional
D5-branes in the sector $+$ and no fractional D5 branes in the sector
$-$. Accordingly, we take $F_+\approx \C^{N_F}$, $F_-=0$ as well as
$\gamma_{0,s}={\bf 1}_{N_F}$ with $N_F=8$.

Finally, we have the massless fields $\phi_1$, $\tilde \phi_1$,
$\phi_2$ and $\tilde \phi_2$ corresponding to $Hom(E_1,F)$,
$Hom(F,E_1)$, $Hom(E_2,F)$ and $Hom(F,E_2)$.  They transform as
follows under the action of the orientifold generator:
\begin{eqnarray}
\phi_1 &\rightarrow & -\gamma_{0,s} \tilde \phi_2^T \gamma_2^{-1} = -\tilde \phi_2^T \\
\phi_2 &\rightarrow &-\gamma_{0,s} \tilde \phi_1^T \gamma_1^{-1} = -\tilde \phi_1^T \\
\tilde \phi_1 &\rightarrow & \gamma_2 \phi_2^T \gamma_{0,s}^{-1} = \phi_2^T \\
\tilde \phi_2 &\rightarrow &-\gamma_1 \phi_1^T \gamma_{0,s} = -\phi_1^T~~.
\end{eqnarray}
This gives the following action and projections for the $\Z_2$
orbifold generator (=the square of the orientifold generator):
\begin{eqnarray}
\phi_1 \rightarrow -\tilde \phi_2^T \rightarrow \phi_1 &\\
\phi_2 \rightarrow -\tilde \phi_1^T \rightarrow -\phi_2 & \Longrightarrow &  \phi_2=0\\
\tilde \phi_1 \rightarrow \phi_2^T \rightarrow -\tilde \phi_1 & \Longrightarrow
&
\tilde \phi_1 =0\\
\tilde \phi_2 \rightarrow -\phi_1^T \rightarrow \tilde \phi_2 & ~~.
\end{eqnarray}
Hence only $\phi:=\tilde \phi_2=-\phi_1^T$ survives, giving eight
fundamentals of the $SU(N)$ gauge group on the wrapped D5-branes.
Thus we recover the massless field content extracted in Appendix
\ref{fractional}.\footnote{The projections are of course equivalent.
  In the fractional brane limit of Appendix B, we extended the $\Z_3$
  orbifold group to $\Z_6$ in order to implement the correct
  projection on the fundamentals as well as fractionality of the
  central D5-brane. In the present appendix, this is implemented by
  the choice $\gamma_{0,a}=0$ in (\ref{gamma0}) .}


\begin{thebibliography}{100}
%
%
\bibitem{vafa} C.~Vafa, ``Superstrings and topological strings at
  large N,'' J.\ Math.\ Phys.\ {\bf 42}, 2798 (2001)
  [arXiv:hep-th/0008142].
%%CITATION = HEP-TH 0008142;%%
%
%

\bibitem{civ} F.~Cachazo, K.~A.~Intriligator and C.~Vafa, ``A large N
  duality via a geometric transition,'' Nucl.\ Phys.\ B {\bf 603}, 3
  (2001) [arXiv:hep-th/0103067].
%%CITATION = HEP-TH 0103067;%%
%
%
\bibitem{Cachazo_Vafa}{ F.~Cachazo, S.~Katz and C.~Vafa, ``Geometric
    transitions and N = 1 quiver theories,'' [arXiv:hep-th/0108120].}
%%CITATION = HEP-TH 0108120;%%
%
%
\bibitem{Cachazo_Vafa_more}{ F.~Cachazo, B.~Fiol, K.~A.~Intriligator,
    S.~Katz and C.~Vafa, ``A geometric unification of dualities,''
    Nucl.\ Phys.\ B {\bf 628}, 3 (2002)[arXiv:hep-th/0110028].}
%%CITATION = HEP-TH 0110028;%%
%
%
\bibitem{DV}{R.~Dijkgraaf and C.~Vafa, ``Matrix models, topological
    strings, and supersymmetric gauge theories,'' Nucl.\ Phys.\ B {\bf
      644}, 3 (2002) [arXiv:hep-th/0206255].}
%%CITATION = HEP-TH 0206255;%%
%
%
\bibitem{DV2}{R.~Dijkgraaf and C.~Vafa, ``On geometry and matrix
    models,'' Nucl.\ Phys.\ B {\bf 644}, 21 (2002)
    [arXiv:hep-th/0207106].}
%%CITATION = HEP-TH 0207106;%%
%
%
\bibitem{DV3}{R.~Dijkgraaf and C.~Vafa, ``A perturbative window into
    non-perturbative physics,'' arXiv:hep-th/0208048.}
%%CITATION = HEP-TH 0208048;%%
%
%
\bibitem{holo}{C.~I.~Lazaroiu, ``Holomorphic matrix models,''JHEP05
    {\bf 05}, 044 (2003)[arXiv:hep-th/0303008].}
%%CITATION = HEP-TH 0303008;%%
%
%
\bibitem{cdsw}{F.~Cachazo, M.~R.~Douglas, N.~Seiberg and E.~Witten,
    ``Chiral rings and anomalies in supersymmetric gauge theory,''
    JHEP {\bf 0212}, 071 (2002)[arXiv:hep-th/0211170].}
%%CITATION = HEP-TH 0211170;%%
%
%
\bibitem{chiral1} K.~Landsteiner, C.~I.~Lazaroiu and R.~Tatar,
  ``Chiral field theories, Konishi anomalies and matrix models,''
  arXiv:hep-th/0307182.
%%CITATION = HEP-TH 0307182;%%
%
%
\bibitem{Karl_chiral}{K.~Landsteiner, E.~Lopez and D.~A.~Lowe,
    ``Duality of chiral N = 1 supersymmetric gauge theories via
    branes,'' JHEP {\bf 9802}, 007 (1998)[arXiv:hep-th/9801002].}
%%CITATION = HEP-TH 9801002;%%
%
%
\bibitem{hana} I.~Brunner, A.~Hanany, A.~Karch and D.~Lust, ``Brane
  dynamics and chiral non-chiral transitions,'' Nucl.\ Phys.\ B {\bf
    528}, 197 (1998) [arXiv:hep-th/9801017].
%%CITATION = HEP-TH 9801017;%%
%
%
\bibitem{kutasov} S.~Elitzur, A.~Giveon, D.~Kutasov and D.~Tsabar,
  ``Branes, orientifolds and chiral gauge theories,'' Nucl.\ Phys.\ B
  {\bf 524}, 251 (1998) [arXiv:hep-th/9801020].
%%CITATION = HEP-TH 9801020;%%
%
%
\bibitem{OT}{K.~h.~Oh and R.~Tatar, ``Duality and confinement in N = 1
    supersymmetric theories from geometric transitions,'', Adv.\
    Theor.\ Math.\ Phys.\ {\bf 6}, 141 (2003) [arXiv:hep-th/0112040].}
%%CITATION = HEP-TH 0112040;%%
%
%
\bibitem{Roemelsberger}{S.~K.~Ashok, R.~Corrado, N.~Halmagyi,
    K.~D.~Kennaway and C.~Romelsberger, ``Unoriented strings, loop
    equations, and N = 1 superpotentials from matrix models,''Phys.\
    Rev.\ D {\bf 67}, 086004 (2003) [arXiv:hep-th/0211291].}
%%CITATION = HEP-TH 0211291;%%
%
%
\bibitem{eot}{J.~D.~Edelstein, K.~Oh, R.~Tatar, ``Orientifold,
    geometric transition and large N duality for SO/Sp gauge
    theories,''JHEP {\bf 0105}, 009 (2001)[arXiv:hep-th/0104037].}
%%CITATION = HEP-TH 0104037;%%
%
%
\bibitem{ookouchi} H.~Fuji and Y.~Ookouchi, ``Confining phase
  superpotentials for SO/Sp gauge theories via geometric transition,''
  JHEP {\bf 0302}, 028 (2003) [arXiv:hep-th/0205301].
%%CITATION = HEP-TH 0205301;%%
%
%
\bibitem{us}{A.~Klemm, K.~Landsteiner, C.~I.~Lazaroiu and I.~Runkel,
    ``Constructing gauge theory geometries from matrix models,'' JHEP
    {\bf 0305}, 066 (2003)[arXiv:hep-th/0303032].}
%%CITATION = HEP-TH 0303032;%%
%
%
\bibitem{or} K.~Landsteiner, C.~I.~Lazaroiu and R.~Tatar,
  ``(Anti)symmetric matter and superpotentials from IIB
  orientifolds,'' arXiv:hep-th/0306236.
%%CITATION = HEP-TH 0306236;%%
%
%
\bibitem{park1} J.~Park, R.~Rabadan and A.~M.~Uranga, ``N = 1 type IIA
  brane configurations, chirality and T-duality,'' Nucl.\ Phys.\ B
  {\bf 570}, 3 (2000) [arXiv:hep-th/9907074].
%%CITATION = HEP-TH 9907074;%%
%
%
\bibitem{park2} J.~Park, R.~Rabadan and A.~M.~Uranga, ``Orientifolding
  the conifold,'' Nucl.\ Phys.\ B {\bf 570}, 38 (2000)
  [arXiv:hep-th/9907086].
%%CITATION = HEP-TH 9907086;%%
%
%
\bibitem{park3} J.~Park, ``M-theory realization of a N = 1
  supersymmetric chiral gauge theory in four dimensions,'' Nucl.\
  Phys.\ B {\bf 550}, 238 (1999) [arXiv:hep-th/9805029].
%%CITATION = HEP-TH 9805029;%%
%
%
\bibitem{lll} K.~Landsteiner, E.~Lopez and D.~A.~Lowe,
  ``Supersymmetric gauge theories from branes and orientifold
  six-planes,'' JHEP {\bf 9807}, 011 (1998) [arXiv:hep-th/9805158].
%%CITATION = HEP-TH 9805158;%%
%
%
\bibitem{Vafa_or}{B.~Acharya, M.~Aganagic, K.~Hori and C.~Vafa,
    ``Orientifolds, mirror symmetry and superpotentials,''
    arXiv:hep-th/0202208.}
%%CITATION = HEP-TH 0202208;%%
%
%
\bibitem{KRS1}
P.~Kraus and M.~Shigemori,
``On the matter of the Dijkgraaf-Vafa conjecture,''
JHEP {\bf 0304}, 052 (2003)
[arXiv:hep-th/0303104].
%%CITATION = HEP-TH 0303104;%%
%
%
\bibitem{Alday}
L.~F.~Alday and M.~Cirafici,
``Effective superpotentials via Konishi anomaly,''
JHEP {\bf 0305}, 041 (2003)
[arXiv:hep-th/0304119].
%%CITATION = HEP-TH 0304119;%%
%
%
\bibitem{KRS}{P.~Kraus, A.~V.~Ryzhov and M.~Shigemori, ``Loop
    equations, matrix models, and N = 1 supersymmetric gauge
    theories,'' arXiv:hep-th/0304138.}
%%CITATION = HEP-TH 0304138;%%
%
\bibitem{BCOV} M.~Bershadsky, S.~Cecotti, H.~Ooguri and C.~Vafa,
  ``Kodaira-Spencer theory of gravity and exact results for quantum
  string amplitudes,'' Commun.\ Math.\ Phys.\ {\bf 165}, 311 (1994)
  [arXiv:hep-th/9309140].
%%CITATION = HEP-TH 9309140;%%
%
%
\bibitem{Csaki}
C.~Csaki, M.~Schmaltz, W.~Skiba and J.~Terning,
``Gauge theories with tensors from branes and orientifolds,''
Phys.\ Rev.\ D {\bf 57}, 7546 (1998)
[arXiv:hep-th/9801207].
%%CITATION = HEP-TH 9801207;%%

%
%
\bibitem{Uranga_probes} A.~M.~Uranga, ``D-brane probes, RR tadpole
  cancellation and K-theory charge,'' Nucl.\ Phys.\ B {\bf 598}, 225
  (2001) [arXiv:hep-th/0011048].
%%CITATION = HEP-TH 0011048;%%
%
%
\bibitem{Douglas_moore} M.~R.~Douglas and G.~W.~Moore, ``D-branes,
  Quivers, and ALE Instantons,'' arXiv:hep-th/9603167.
%%CITATION = HEP-TH 9603167;%%

\end{thebibliography}
\end{document}